\documentclass[notitlepage,nofootinbib,eqsecnum]{revtex4-1}
\usepackage{amsmath,bm,graphicx,caption,amssymb,float}
\usepackage[latin2]{inputenc}

\usepackage[colorlinks=true,backref=false, linktocpage=true,
citecolor=blue,urlcolor=blue,linkcolor=blue,pdfpagemode=UseOutlines,
pdfstartview=FitH,bookmarksopen]{hyperref}
\usepackage{bookmark}
\bookmarksetup{
  numbered, 
  open,
}

\newcommand{\Ca}{A}
\newcommand\VE{{V}}
\newcommand\hb{{\bar h}}
\newcommand\rb{{\bar r}}

\DeclareMathAlphabet{\mathpzc}{OT1}{pzc}{m}{it}

\graphicspath{{figures/}}

\begin{document}

\title{Universality of the critical point mapping between Ising model and QCD at small quark mass}

\author{Maneesha Sushama Pradeep}
\author{Mikhail Stephanov}

\affiliation{Department of Physics, University of Illinois, Chicago,
  IL 60607, USA}

\date{\today}

\begin{abstract}
  The universality of the QCD equation of state near the critical
  point is expressed by mapping pressure as a function of temperature
  $T$ and baryon chemical potential $\mu$ in QCD to Gibbs free energy
  as a function of reduced temperature $r$ and magnetic field $h$ in
  the Ising model. The mapping parameters are, in general, not
  universal, i.e., determined by details of the microscopic dynamics,
  rather than by symmetries and long-distance dynamics. In this paper
  we point out that in the limit of small quark masses, when the
  critical point is close to the tricritical point, the mapping
  parameters show {\em universal} dependence on the quark mass
  $m_q$. In particular, the angle between the $r=0$ and $h=0$ lines in
  the $(\mu,T)$ plane vanishes as $m_q^{2/5}$. We discuss possible
  phenomenological consequences of these findings.
\end{abstract}

\maketitle

\section{Introduction}
\label{sec:introduction}

Mapping QCD phase diagram is one of the fundamental goals of heavy-ion
collision experiments as well as lattice gauge theory
computations. The QCD critical point is one of the crucial features of
the phase diagram~\cite{Stephanov:2004wx}. The position and even the
existence of this point is still an open question. The potential for
discovery of the QCD critical point is one of the major motivations
for the ongoing Beam Energy Scan program at RHIC as well as future
heavy-ion collision experiments~\cite{Luo:2017faz}.

The straightforward reliable determination of the location of the critical
point by lattice QCD computations~\cite{Ding:2015ona} is impeded by the notorious
sign problem. However, even in the absence of such a first-principle
calculation one can predict some specific properties of QCD in the
vicinity of the critical point. These properties follow from the
universality of the critical behavior. In this paper we shall focus on
static thermodynamic properties which are described by the equation of
state. Besides having fundamental significance, the QCD
equation of state is a crucial input in hydrodynamic calculations aimed
at describing the heavy-ion collisions and identifying the signatures
of the critical point.

The universality of static critical phenomena allows us to predict the
leading singular behavior of thermodynamic functions, such as pressure
$P(\mu,T)$ on temperature and chemical potential. The leading singular
contribution to the QCD equation of state is essentially the same as 
the singular part of the equation of state of the Ising model with
$\mu$ and $T$ in QCD mapped onto (reduced) temperature $r=T-T_c$ and
ordering (magnetic) field $h$ of the Ising model. The parameters of
the mapping are not universal and are generally treated as unknown
parameters. 

In this paper we shall investigate the properties of this mapping in
order to constrain or determine a reasonable domain for the values of
the unknown mapping parameters. Our main finding follows from the fact
that, due to the smallness of the (light) quark mass $m_q$ the
{\em critical} point is close, in parameter space, to the {\em
  tricritical} point \cite{Stephanov:1998dy} -- the point separating
the second and first-order finite temperature chiral restoration
transition.\footnote{These considerations would also apply, {\it
    mutatis mutandis}, to the tricritical point separating the second
  and first-order transitions as a function of the strange quark mass
  \cite{Rajagopal:1992qz,Gavin:1993yk}, instead of the baryon chemical
  potential.}  Thermodynamics near the tricritical point is also
universal, albeit the universality class is different from the one of
the Ising model. We point out that certain properties of the
$(\mu,T)/(h,r)$ {\em mapping} near the critical point are universal in
the limit of small quark masses due to the proximity of the
tricritical point. The mapping becomes singular in a specific
way. Most importantly, we observe that the slopes of the $r=0$ and
$h=0$ lines in the $(\mu,T)$ plane become increasingly aligned near the
critical point, with the slope difference vanishing with a specific
power of the quark mass: $m_q^{2/5}$.

The paper is organized as follows: In Section~\ref{sec:mapping-qcd-3d}
we describe the mapping between QCD and Ising critical equations of
state, set notations and derive useful relations which allow us to
determine the mapping parameters from a given equation of state. In
Section~\ref{sec:mean-field-eos} we describe how to determine the
non-universal mapping parameters in a generic Ginzburg-Landau, or
mean-field, theory of the critical point. In
Section~\ref{sec:effective_tricritical} we apply the results of
Sections~\ref{sec:mapping-qcd-3d} and \ref{sec:mean-field-eos} to
determine mapping parameters in a special case where a critical point
is close to a tricritical point, which is also described by
Ginzburg-Landau theory. We show that the mapping becomes singular,
i.e., the slopes of $h=0$ and $r=0$ lines converge with the difference
vanishing as $m_q^{2/5}$. In Section~\ref{sec:rmm} we use Random Matrix
Model of QCD to illustrate our results and estimate the values of
mapping parameters for a physical value of quark
mass. In Section~\ref{sec:corrections} we investigate the effect of
fluctuations, i.e., go beyond mean-field approximation using epsilon
expansion. We show that the main conclusion -- convergence of the
slopes with difference of order $m_q^{2/5}$ is robust at least to
two-loop order. We conclude in
Section~\ref{sec:concl-disc} and discuss possible phenomenological implications.

\section{Mapping QCD to 3D Ising model}
\label{sec:mapping-qcd-3d}
The universality of the critical phenomena is a consequence of the
fact that these phenomena are associated with the behavior (such as
response or fluctuations) of the critical systems at scales much
longer than the microscopic scales (e.g., interparticle
distances). Such response is nontrivial in critical systems because of
the  large (divergent at the critical point) correlation
length. As a result, critical fluctuations and response can be
described by a field theory which becomes conformal at the critical
point. Microscopically different theories which have the same
conformal fixed point in the infrared can therefore be mapped onto
each other. For example, all liquid-gas critical points can be mapped
onto the critical point of the Ising model because all flow to the same
infrared fixed point described by the one-component $\phi^4$ theory at
the Wilson-Fischer fixed point. The universality class corresponding
to this conformal fixed point is the most
ubiquitous in Nature\footnote{This fixed point does not require any
  continuous symmetries which are typically necessary to maintain degeneracy
  between {\em multiple\/} components of the order-parameter field as
  is the case, for example, in the $O(3)$ Heisenberg ferromagnet.}
and the QCD critical point, if it exists, belongs to it.

There are two relevant parameters in the $\phi^4$ theory which need to
be tuned to zero to reach the critical point: these are the
coefficients of the two relevant operators, $\phi$ and $\phi^2$. In
the Ising model, due to the $Z_2$ symmetry $\phi\to-\phi$, they map
directly onto the ordering (magnetic) field $h$ and reduced
temperature $r=T-T_c$, with no mixing. In QCD, or for a generic
liquid-gas critical point, the parameters which need to be tuned are
temperature $T$ and chemical potential $\mu$ and neither of them have
any particular relation to the $Z_2$ symmetry (in fact, there is no
$Z_2$ symmetry except in the scaling regime near the critical
point). Therefore one should expect a generic mapping $h(\mu,T)$ and
$r(\mu,T)$.

\subsection{Definition of mapping parameters}
\label{sec:defin-mapp-param}

The universality is expressed by the relation of the partition
functions of QCD and the Ising model near the critical point if
expressed in terms of variables $r$ and $h$. Since the pressure in QCD
and Gibbs free energy in the Ising model are both proportional to the
logarithms of the respective partition functions one can write:
\begin{equation}
\label{pG}
P_{\text{sing}}(\mu,T)=-\Ca G(r(\mu,T), h(\mu,T))\,,
\end{equation}
where $P_{\text{sing}}$ is the leading singular term in the QCD
pressure at the critical point and $G$ is the singular term in the
Gibbs free energy of the Ising model, or $\phi^4$ theory. The
relation~(\ref{pG}) and the corresponding $(\mu,T)/(h,r)$ mapping was
introduced by Rehr and Mermin and termed ``revised
scaling''\footnote{The original version of scaling equation of state
  by Widom in Ref.~\cite{Widom} mapped $r$ to $T-T_c$ directly,
  without allowing for mixing with~$h$, which did not account for the
  asymmetry on the coexistence line found in liquid-gas transitions
  (e.g., discontinuity of susceptibility). This original scaling
  corresponds, in the notations used in the present paper, to
  $\alpha_2=0$. } in Ref.\cite{Rehr:1973zz}. In relativistic field
theories it has been studied in the context of QCD, e.g., in
Refs.\cite{Karsch:2001nf,Hatta:2002sj,Nonaka:2004pg,Bluhm:2006av,Parotto:2018pwx,Akamatsu:2018vjr},
and, earlier, in the context of the electroweak transition in
Ref.\cite{Rummukainen:1998as}.

\begin{figure}[h]
\begin{center}
  \includegraphics[scale=0.6]{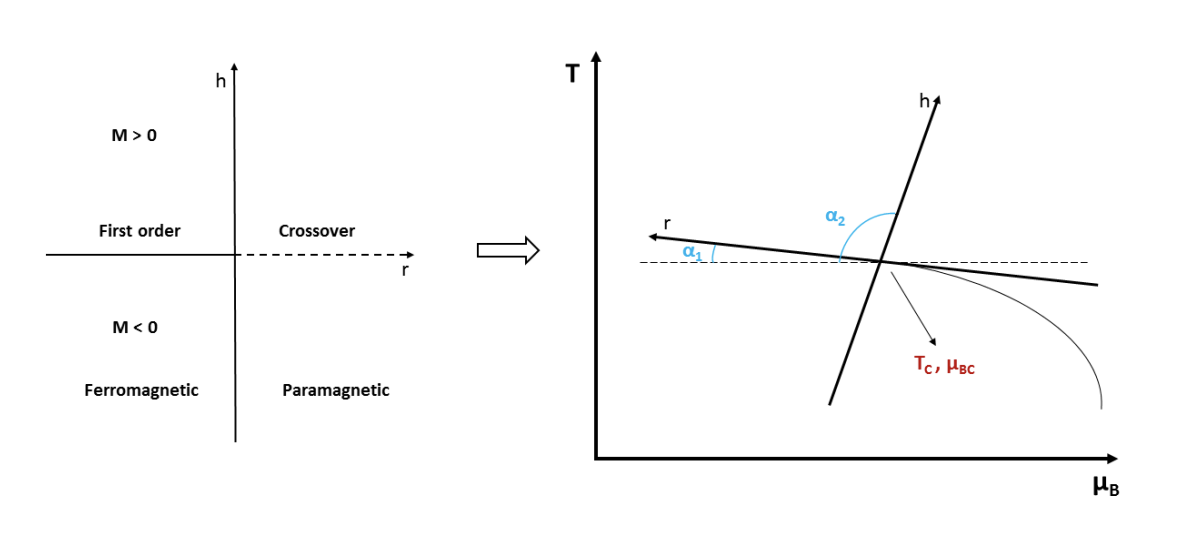}
  \caption{The mapping between QCD and Ising variables given by
    Eq.~(\ref{map0}). Note that, since the sign of $h$ is a matter of
    convention, the mappings with $\alpha_2$ and $\alpha_2\pm\pi$ are
    essentially equivalent. The figure is taken from
    Ref.~\cite{Parotto:2018pwx}.}
\label{fig:paolo_map}
  \end{center}
  \end{figure}

 Let
$(\mu_c,T_c)$ be the location of the QCD critical point on the $(\mu,T)$
plane. To describe the leading singularity it is sufficient to
linearize the mapping functions $h(\mu,T)$ and $r(\mu,T)$ in
$\Delta T=T-T_c$ and $\Delta \mu=\mu-\mu_c$. We follow the convention
for the coefficients of the linear mapping introduced in
Ref.~\cite{Parotto:2018pwx}:
\begin{eqnarray}\nonumber
h(\mu,T)&=&
h_T\Delta T + h_\mu\Delta\mu
=-\frac{\cos \alpha_1 \Delta T+\sin \alpha_1 \Delta \mu}{w T_c \sin(\alpha_1-\alpha_2)};\\  \label{map0}
 r(\mu,T)&=&
r_T\Delta T + r_\mu\Delta\mu
=\frac{\cos \alpha_2 \Delta T+\sin \alpha_2 \Delta \mu}{\rho w T_c\sin(\alpha_1-\alpha_2)}\,,
 \end{eqnarray}
where we denoted by a subscript $T$ or $\mu$ the partial derivative with
respect to the corresponding variable, e.g., $h_T\equiv \partial
h/\partial T$ at fixed $\mu$. Additional parameters $w$ and $\rho$ provide
absolute and relative normalization of $h$ and $r$ setting  the
size and shape of the critical region (see Appendix~\ref{sec:size-critical-region}).
The angles $\alpha_1$ and $\alpha_2$ describe the slopes of the lines $h=0$ ($r$ axis) and $r=0$ ($h$ axis) on the
$(\mu,T)$ plane, as shown in Fig.~\ref{fig:paolo_map}:
\begin{eqnarray}
\label{t1}
\left(\frac{dT}{d\mu}\right)_{h=0}&=&
=-\frac{h_\mu}{h_T}
=-\tan\alpha_1;\\ \label{t2}
\left(\frac{dT}{d\mu}\right)_{r=0}&=&
=-\frac{r_\mu}{r_T}
=-\tan\alpha_2\,.
\end{eqnarray}

An important property of the most singular part of the Ising Gibbs free energy
is scaling:
\begin{equation}
  \label{eq:Gscaling1}
  G(\lambda r, \lambda^{\beta\delta} h) = \lambda^{\beta(\delta+1)} G (r,h)
\end{equation}
with well-known critical exponents $\beta$ and $\delta$. Another
important property is the $Z_2$ symmetry:
\begin{equation}
  \label{eq:Gscaling2}
  G( r, - h) = G (r, h)\,.
\end{equation}

Eqs.~(\ref{eq:Gscaling1}) and (\ref{eq:Gscaling2}) together imply that the function $G$ can be written in terms of an even function $g$ of one variable only:
\begin{equation}
  \label{eq:g}
  G(r,h) = r^{\beta(\delta+1)} g(hr^{-\beta\delta})\,.
\end{equation}
The universal scaling function $g$ is multivalued. In the complex
plane of its argument $x\equiv hr^{-\beta\delta}$ the primary Riemann
sheet describes the equation of state at high temperatures $T>T_c$,
i.e., $r>0$.  We shall denote the value of $g$ on this sheet as
$g_+(x)$.  On the primary (high temperature) sheet the function is
analytic at $x=0$. The closest singularities are on the imaginary $x$
axis and are known as Lee-Yang edge singularities~\cite{Lee:1952ig,Fisher:1978pf}
(see also recent discussions in Refs.\cite{Fonseca:2001dc,An:2017brc}). A
secondary Riemann sheet describes the low temperature phase, $r<0$. We
shall denote the value of $g$ on this sheet as $g_-(x)$.

\subsection{Relation between mapping parameters and derivatives of
  pressure}
\label{sec:relat-betw-mapp}

Given an equation of state $P(\mu,T)$ one should be able to determine
the mapping parameters. In this Subsection, for further applications, we shall
derive expressions which can be used to do that.

We shall take all derivatives of pressure below on the crossover line,
i.e., at $h=0$ for $r>0$ and keep only the most singular terms.
In this manuscript, a subscript with respect to $\mu$ or
$T$ implies differentiation with respect to that variable when the
other is kept fixed. Below, we also use indices $X$  and $Y$ to
represent either $T$ or $\mu$. We find, at $h=0$:
 \begin{eqnarray} 
\label{pTmu}
P_{XY}&=&-\Ca h_{X}h_{Y}r^{\beta-\beta\delta}g_{+}^{''}(0)+\ldots;\\ 
\label{pmumumu1}
P_{XXX}&=&3\Ca(\beta\delta-\beta)
           h_{X}^2r_{X}r^{\beta-\beta\delta-1}g_{+}^{''}(0)+\ldots; \\ 
\label{pmumuTT}
P_{XX YY}&=&-\Ca
             h_{X}^2h_{Y}^2r^{\beta-3\beta\delta}g_{+}^{''''}(0)+\ldots. 
\end{eqnarray} 
The dots represent the terms which are subleading to the terms
explicitly written in the limit $r\rightarrow 0$. From the above
equations, it is easy to see that
\begin{eqnarray}
  \label{eq:hx}
h_X &=& \lim_{\substack{h=0 \\ r\rightarrow0^{+}}}
\left(\frac{P_{XX}r^{\beta\delta-\beta}}{-Ag_+''(0)}
\right)^{1/2};\\
\label{eq:rx}
r_X &=& \lim_{\substack{h=0 \\ r\rightarrow0^{+}}}
\frac{P_{XXX}r}{3(\beta-\beta\delta)P_{XX}}\,.
\end{eqnarray}
Therefore:
\begin{alignat}{2}
\label{h_sl}
\tan \alpha_1&=\frac{h_{\mu}}{h_T}&&=\lim_{\substack{h=0 \\ r\rightarrow0^{+}}}\frac{P_{\mu\mu}}{P_{\mu T}}\,;\\
\label{r_sl}
\tan \alpha_2&=\frac{r_{\mu}}{r_T}&&=\lim_{\substack{h=0 \\ r\rightarrow0^{+}}}\frac{P_{\mu\mu\mu}}{P_{TTT}}\frac{P_{TT}}{P_{\mu\mu}}\,.
\end{alignat}

Eq.~(\ref{h_sl}) simply means, in particular, that the slope of the
contour of critical number density ($n=P_\mu$) at the critical point is equal to
the slope of $h=0$. Eq.~(\ref{r_sl}) relates the slope of $r=0$ to
the ratios of third and second derivatives of pressure evaluated along
the cross-over line. These equations could be compared and contrasted
with the expressions obtained by Rehr and  Mermin
in Ref.~\cite{Rehr:1973zz} using the discontinuities of the 
derivatives of pressure along the first-order line.

Similarly, the parameters $\rho$ and $w$ in the mapping can also be
related to pressure derivatives. In order to do that we also need an
expression for $r$:
\begin{eqnarray}
\label{r1}
r&=&\left(-\frac{g^{''}(0)^2 \Ca\,P_{\mu\mu TT}}{g^{''''}(0)P_{\mu\mu}P_{TT}}\right)^{-\frac{1}{\beta(\delta+1)}}\,.
\end{eqnarray}
Using that expression in Eqs.~(\ref{eq:hx}) and~(\ref{eq:rx}) we can
obtain $\rho$ and $w$ by substituting $h_X$ and $r_X$ into the
following expressions:
\begin{eqnarray}\label{eq:rho-hr}
\rho&=&\sqrt{\frac{h_\mu^2+h_T^2}{r_{\mu}^2+r_T^2}};\\
\label{eq:w-hr}
w T_c &=& \frac{\sqrt{r_\mu^2+r_T^2}}{|r_Th_\mu-r_\mu h_T|}.
\end{eqnarray}

The normalization convention in Ref.~\cite{Parotto:2018pwx} which we
follow corresponds to  $\Ca=T_c^4/2$. Obviously, the
angles $\alpha_1$ and $\alpha_2$ do not depend on this normalization
  whereas $\rho$ and $w$ do.
To fix the normalization of $h$ and $r$ we follow the standard
convention, also used in Ref.~\cite{Parotto:2018pwx}:
\begin{equation}
  \label{norm_Ising}
  g_{-}'(0^{+})=-(-1)^\beta \, , \qquad \lim_{x\rightarrow\infty^{+}} x^{-1/\delta} g_{\pm}'(x)=-1\,.
\end{equation}
	
Using equations in this subsection we can determine $\alpha_1$,
$\alpha_2$, $\rho$ and $w$ if the pressure is known as a function of
$\mu$ and $T$. It should be mentioned that we chose one among many
ways of expressing $\alpha_1$, $\alpha_2$, $\rho$ and $w$ in terms of
ratios of pressure derivatives. Our choice was guided by the desire to
obtain expressions which treat $T$ and $\mu$ variables most
symmetrically.

\section{Mean-field equation of state}
\label{sec:mean-field-eos}

\subsection{Symmetry and scaling in mean-field theory}
\label{sec:scaling-mean-field}

In the mean-field description of the critical point equation of state,
pressure can be expressed as the minimum of the Ginzburg-Landau
potential as a function of the order parameter $\phi$:
\begin{equation}
 \label{Omega1}
 P(\mu,T)=-\Ca\min_{\phi}\Omega(\phi,\mu,T)\,.
\end{equation}
Let us make a simple but very useful observation: by a change of
variable $\phi\to f(\phi)$ one can obtain a family of potentials
$\hat\Omega(\phi)$ obeying $\hat\Omega(\phi)=\Omega(f(\phi))$ each of
which gives the same pressure. We shall refer to this property as
reparametrization invariance.
 
Close to the critical point, $\Omega$ can be expanded around the
critical value of $\phi$ (chosen to be $\phi=0$):
\begin{eqnarray}
\label{Oseries1}
  \Omega(\phi,\mu,T)=\Omega_0
  -h \phi+\frac{r}{2} \phi^2+\frac{u}{4} \phi^4 + \ldots\,,
\end{eqnarray}
where we eliminated cubic term $\phi^3$ by a shift of variable $\phi$
(such an operator or term is called redundant in renormalization group
terminology). Parameters $\Omega_0$, $h$, $r$ and $u$ are analytic
functions of $\mu$ and $T$. The critical point is located at $h=0$ and
$r=0$ (with $u>0$).  If we truncate the expansion at order $\phi^4$ as
in Eq.~(\ref{Oseries1}) the $\phi$-dependent part, $\Omega-\Omega_0$,
possesses two important properties. 
The first is the $Z_2$ symmetry:
\begin{equation}
  \label{eq:Z2}
  \phi\to-\phi , \quad h\to-h,  \quad r\to r\,. 
\end{equation}
The second is scaling:
\begin{equation}
  \label{eq:mf-scaling}
  \phi\sim r^{1/2}, \quad h\sim r^{3/2}, \quad \Omega-\Omega_0\sim r^2\,.
\end{equation}
This corresponds to the scaling of the Gibbs free energy $G$ in
Eq.~(\ref{eq:Gscaling1}) with mean-field exponents $\beta=1/2$ and $\beta\delta=3/2$.

One could be tempted to expand the coefficients $h$ and $r$ in
Eq.~(\ref{Oseries1}) to linear order in $\Delta T$ and $\Delta\mu$ and
identify the mixing parameters $\alpha_1$, $\alpha_2$, etc., by using
Eq.~(\ref{map0}). This, however, is not entirely correct as it would
ignore the fact that the mixing of $h$ and $r$ described by
Eq.~(\ref{map0}) necessarily violates scaling, since $h\sim r^{3/2}$
and $r$ have different scaling exponents. Therefore, we need also to look
at the omitted terms which violate scaling in Eq.~(\ref{Oseries1}), or more
precisely, provide corrections to scaling of relative order $h/r\sim
r^{1/2}$ (i.e., $r^{\beta\delta -1}$). 
Furthermore, mixing of $h$ and
$r$ also violates $Z_2$ symmetry in Eq.~(\ref{eq:Z2}), i.e., we need
also to look at omitted $Z_2$ breaking terms in Eq.~(\ref{Oseries1}).

Since $\phi\sim r^{1/2}$, omitted higher order terms in
Eq.~(\ref{Oseries1}) represent corrections to scaling. The leading
correction is due to the $\phi^5$ term. Because in mean-field theory
this term is smaller by exactly a factor of $r^{1/2}$ compared to the terms in
Eq.~(\ref{Oseries1}), and also because it violates the $Z_2$ symmetry
in Eq.~(\ref{eq:Z2}) (being odd), this term will affect the mixing of
$h$ and $r$.

\subsection{The effect of the $\phi^5$ term}
\label{sec:effect-phi5-term}

Let us denote the coupling of the $\phi^5$ term by $v u$, i.e.,
\begin{eqnarray}
\label{Oseries}
\Omega=\Omega_0-\hb \phi+\frac{1}{2}\rb \phi^2+ \frac{u}{4} \phi^4+v u\phi^{5}+O(\phi^6),
\end{eqnarray}
where we also changed the notation for the coefficients of the $\phi$
and $\phi^2$ terms in anticipation of them being different from
$h$ and $r$ in Eq.~(\ref{map0}).

To understand the effect of the $\phi^5$ term on the mixing of $h$ and
$r$ we can use reparametrization invariance of pressure to change the
variable $\phi$ in such a way as to eliminate the $\phi^5$ term from
$\Omega$. This can be achieved by the following transformation:
\begin{equation} \label{m_as1}
\phi \to \phi+v\left(\frac{\rb}{u}- \phi^2\right)\,,
\end{equation}
which eliminates $\phi^5$ and as well as $\phi^3$ term at order up to $r^{5/2}$:
\begin{eqnarray}
\label{Oseries6}
  \Omega=\left(\Omega_0-\frac{v\hb\rb}{u}\right)
  -\left( \hb - \frac{v\rb^2}{u}\right)\phi
  +\left(\frac{\rb}{2} + v\hb \right) \phi^2
  +\frac{u}{4} \phi^4 + \mathcal O(\phi^6,r^{3})\,,
\end{eqnarray}
where we kept only terms up to order $r^{5/2}$, since we are interested in
the leading correction to scaling.
From Eq.~(\ref{Oseries6}) we can now read off the parameters $h$ and
$r$:
\begin{eqnarray}
h&=&u^{-1/4}\left(\hb-\frac{v\rb^2}{u}\right)
=u^{-1/4}\hb + \mathcal O(\rb^2), \label{h_as1}\\ 
r&=&u^{-1/2}\left(\rb+2v\hb\right),           \label{r_as1}
\end{eqnarray}
which match Eq.~(\ref{Oseries6}) onto a mean-field potential without leading asymmetric
($Z_2$-breaking, non-Ising) corrections to scaling. The additional
rescaling $\phi\to u^{-1/4}\phi$ was applied to bring 
the potential to the canonical form:\footnote{The rescaling does not affect the slopes of $h=0$ or $r=0$ (angles
$\alpha_1$ and $\alpha_2$), but needs to be taken into account when
calculating $\rho$ and $w$.
}
\begin{equation}
  \label{eq:Omega-canon}
  \Omega = -h\phi + \frac{r}2\phi^2 + \frac14\phi^4.
\end{equation}
The scaling function $g(x)$ corresponding to this potential via
$G=\min_\phi\Omega=r^{2}g(hr^{-3/2})$ (see Eq.~(\ref{eq:g})) satisfies 
\begin{equation}
x+g'(x)+g'^3(x)=0, \label{eq:g'}
\end{equation}
which agrees with the normalization in Eq.~(\ref{norm_Ising}).
Therefore, parameters $h$ and $r$ in Eqs.~(\ref{h_as1})
and~(\ref{r_as1}) are the parameters which appear in the mapping
equations (\ref{map0}).

Note that the main effect of the asymmetric corrections to scaling is
to modify $\rb$ in Eq.~(\ref{r_as1}) by a term {\em linear} in $\hb$,
which has direct effect on the angle $\alpha_2$ determining the slope
of the $r=0$ axis. The slope of the $h=0$ axis is not affected as the
shift of $\hb$ in Eq.~(\ref{h_as1}) is quadratic in $\rb$.

\subsection{Direct relation to derivatives of the potential}
\label{sec:relat-deriv}

It is also useful to relate mapping parameters
$h_X$ and $r_X$, where $X=T$ or $\mu$, directly to the Ginzburg-Landau
potential $\Omega$. The relation can be obtained straightforwardly from
Eqs.~(\ref{pTmu}-\ref{pmumuTT}) using
\begin{eqnarray}\label{pmumu}
P_{XX}&=&-\Omega_{XX}+\Omega_{X\phi}^2\Omega_{\phi\phi}^{-1}\,,\\ \label{pmumumu}
P_{XXX}&=&-\Omega_{XXX}-3\Omega_{\phi\phi}^{-2}\Omega_{X\phi}(\Omega_{X\phi}\Omega_{X\phi\phi}-\Omega_{\phi\phi}\Omega_{XX\phi})+\Omega_{\phi
           X}^3\Omega_{\phi\phi}^{-3}\Omega_{\phi\phi\phi}\,.
\end{eqnarray}
To simplify the expressions we shall first consider potential
$\hat\Omega$ obtained from $\Omega$ by bringing it into the ``Ising''
form in Eq.~(\ref{Oseries1}) with no $\phi^3$ or $\phi^5$ terms (up to
order $r^{5/2}$). We showed that this can be always achieved by a
reparametrization as in Eq.~(\ref{m_as1}), Eq.~(\ref{Oseries6}). In
this case, $\hat\Omega_{\phi\phi\phi}=0$ on the $h=0$ line along which
we take the limits in Eqs.~(\ref{eq:hx}),~(\ref{eq:rx}) and
expressions simplify:
  \begin{eqnarray}
  \label{alpha1}
\tan \alpha_1 &=& \frac{\hat\Omega_{\phi\mu}}{\hat\Omega_{\phi T}}\,;\\
\label{alpha2}
 \tan \alpha_2 &=& \frac{\hat\Omega_{\phi\phi\mu}}{\hat\Omega_{\phi\phi T}}\,;\\
 \label{rho}
\rho&=&\left(\frac{\hat\Omega_{\phi\phi\phi\phi}}{6}\right)^{1/4}\sqrt{\frac{\hat\Omega_{\phi\mu}^2+\hat\Omega_{\phi T}^2}{\hat\Omega_{\phi\phi\mu}^2+\hat\Omega_{\phi\phi T}^2}}\,;\\
\label{w}
wT_c&=&\left(\frac{\hat\Omega_{\phi\phi\phi\phi}}{6}\right)^{1/4}\frac{\sqrt{\hat\Omega_{\phi\phi\mu}^2+\hat\Omega_{\phi\phi T}^2}}{|\hat\Omega_{\phi\mu}\hat\Omega_{\phi\phi T}-\hat\Omega_{\phi T}\hat\Omega_{\phi\phi\mu}|}\,.
\end{eqnarray}
Note that in the mean-field theory these expressions are analytic at the
critical point and can be simply evaluated at the critical point
without taking a limit. This is in contrast to Eqs.~(\ref{eq:hx})
and~(\ref{eq:rx}) where the derivatives of pressure are singular and
a careful limit has to be taken to cancel singularities.

One can then generalize these expressions to arbitrary
  potential ($\Omega_{\phi\phi\phi}\neq 0$ at $h=0$) by observing that combinations
  \begin{equation}
\Omega_{\phi X}, 
\quad
\Omega_{\phi\phi X}
-\frac{\Omega_{\phi\phi\phi\phi\phi}\Omega_{\phi
    X}}{10\Omega_{\phi\phi\phi\phi}},
\quad\mbox{and}\quad
\Omega_{\phi\phi\phi\phi},
\label{eq:covariant}
\end{equation}
are reparametrization ``covariant'' to leading
order in $r$ in the sense that under $\phi\to f(\phi)$ they transform
multiplicatively by factors $f'$, $(f')^2$ and $(f')^4$,
respectively. Thus, we can drop `hats' and replace
 \begin{equation}
\hat\Omega_{\phi\phi X} \to \Omega_{\phi\phi X}
-\frac{\Omega_{\phi\phi\phi\phi\phi}\Omega_{\phi
    X}}{10\Omega_{\phi\phi\phi\phi}}
\label{eq:O''-subs}
\end{equation}
in Eqs.~(\ref{alpha1}-\ref{w})  to obtain general formulas applicable to
any potential. Note that the last term in Eq.~(\ref{eq:O''-subs})
corresponds to the last term in Eq.~(\ref{r_as1}) describing the
mixing of $r$ and $h$ due to the $\phi^5$ term.

\section{Critical point near a tricritical point}
\label{sec:effective_tricritical}	
	
A tricritical point arises in many systems where the order of the
finite-temperature transition from broken to restored symmetry phase
depends on an additional thermodynamic parameter, such as pressure or
chemical potential. The point where the order of the transition
changes from second to first is a tricritical point. There are reasons
to believe QCD to be one of the examples of such a theory
\cite{Stephanov:1998dy,Stephanov:2004wx}.  A nonzero value of a
parameter which breaks spontaneously broken symmetry explicitly (quark
mass in QCD) removes the second order phase transition and replaces it
with analytic crossover, while the first order transition then ends at
a critical point.

We shall apply mean-field theory near the tricritical point. The
potential needed to describe the change from a first to second order
transition needs to include a $\Phi^6$ term which becomes marginal in
$d=3$. Therefore, mean field theory should be applicable in $d=3$ if
one is willing, as we are, to neglect small logarithmic corrections to
scaling.\footnote{For example, if these corrections have
  negligible consequences for applications, such as heavy-ion
  collisions or lattice QCD simulations. To be rigorous, we can also
  formally consider $d>3$. In fact, our analysis near the {\em critical} point
  is constrained by an even stronger condition, since the upper
  critical dimension in this case is $d=4$ and, in practice, we work in
  $d=4-\epsilon>3$ when we study the effects of fluctuations in Section~\ref{sec:corrections}. }

As in Section~\ref{sec:mean-field-eos} we want to express the pressure
as a minimum of the Ginzburg-Landau potential $\Omega$. We can do that
using 
 the Legendre transform of pressure $P$ with respect to $m_q$:
	\begin{equation}
	V(\Phi,\mu,T)= -P(\mu,T,m_q(\Phi))+m_q(\Phi)\Phi\,,
	\end{equation}
where $m_{q}(\Phi)$ is the solution of
\begin{equation}
{\partial P}/{\partial m_q}=\Phi\label{eq:dPdM},
\end{equation}
which means $\Phi$ is the chiral condensate (times $N_f$ -- the number
of light quarks).

It is easy to see that the potential $\Omega$ defined as 
\begin{equation}
\Ca\Omega(\Phi,\mu,T,m_q)=V(\Phi,\mu,T)-m_q \Phi
\end{equation}
is related to pressure by
\begin{equation}
	\label{pminV}
	P(\mu,T,m_q)=-\Ca\min_{\Phi} \Omega(\Phi,\mu,T,m_q)
	\end{equation}
where we chose the normalization constant $A$ to match Eq.~(\ref{Omega1}).

The potential $V$ has to be symmetric
under $\Phi\rightarrow-\Phi$ (this is a discrete
subgroup of the continuous chiral symmetry) and  to describe a
tricritical point we need terms up to $\Phi^6$.
Expanding $V$ we find:
\begin{equation}\label{eq:V=abc}
 V(\Phi,\mu,T)=V_0+\frac{a}{2}\Phi^2+\frac{b}{4}\Phi^4+\frac{c}{6}\Phi^6
        +\ldots\,,
 \end{equation}
where $a$, $b$ and $c$ are functions of $T$ and $\mu$.  The
tricritical point occurs when $a=b=0$ with $c>0$. If we truncate $V$ at
order $\Phi^6$ as in Eq.~(\ref{eq:V=abc}) the $\Phi$-dependent part of
$V$ and $\Omega$,
has the following scaling property:
\begin{equation}
  \label{eq:3cr-scaling}
  \Phi\sim a^{1/4},\quad b\sim a^{1/2}, \quad m_q\sim a^{5/4}\quad V-V_0\sim a^{3/2}\,.
\end{equation}
  
The minimum value of $\Omega$ in Eq.~(\ref{pminV}) is achieved at $\Phi$ satisfying, to
lowest order in $a\to0$, 
\begin{equation}
  \label{eq:eos3}
  m_q = \frac{\partial \VE}{\partial\Phi} = a\Phi + b\Phi^3 + c \Phi^5 \,.
\end{equation}
At nonzero $m_q$ the critical point occurs when both second and third
derivatives of $\Omega$ vanish at the minimum given by
Eq.~(\ref{eq:eos3}). I.e.,
\begin{equation}
  \label{eq:dmdphi}
  \frac{\partial^2\VE}{\partial\Phi^2} = \frac{\partial
    m_q}{\partial\Phi} = a + 3b\Phi^2 + 5c\Phi^4 =0.
\end{equation}
and
\begin{equation}
  \label{eq:d3Omega}
   \frac{\partial^3\VE}{\partial\Phi^3}
   = 6b\Phi + 20c\Phi^3 =0.
\end{equation}
 Eqs.~(\ref{eq:eos3}),~(\ref{eq:dmdphi}) and~(\ref{eq:d3Omega}), can be solved simultaneously to find the critical values of $\Phi$, $a$
and $b$ for a given $m_q$:
\begin{equation}
  \label{eq:phiab}
  \Phi_c = \left( \frac{3m_q}{8c}\right)^{1/5},\quad a_c =
  5c\Phi_c^4,\quad b_c = - \frac{10c}{3}\Phi_c^2.
\end{equation}
As a function of $m_q$, the trajectory 
$(m_q,a_c(m_q),b_c(m_q))$ corresponds to the line of critical points
on the edges of ``wings'' -- coexistence
surfaces in the $m_q$, $T$, $\mu$ phase diagram (see, e.g.,
Fig.~\ref{fig:pd-3d-rmm} for illustration). Note that
critical values of parameters in Eq.~(\ref{eq:phiab})
 scale as $\Phi_c\sim
m_q^{1/5}, a_c\sim m_q^{4/5}$ and $b_c\sim m_q^{2/5}$ consistent with the
scaling in Eq.~(\ref{eq:3cr-scaling}). We can now expand $\Omega$ around that solution: 
\begin{multline}
  \label{V:-expanded}
\Ca\, \Omega(\Phi;\mu,T,m_q)= \Ca\,\Omega(\Phi_c;T_c,\mu_c,m_q) +  (\Delta a \Phi_c + \Delta b \Phi_c^3)\phi+
 \frac{1}{2}  (\Delta a  + 3\Delta b \Phi_c^2) \phi^2
  +\frac{1}{4} \left( \frac{20c\Phi_c^2}{3} + \Delta b \right) \phi^4\\
+ \Delta b\Phi_c \phi^3 
  + c\Phi_c \phi^5 
 +\frac{c}{6} \phi^6
\,,
\end{multline}
where $\Delta a = a-a_c$, $\Delta b = b - b_c$ and
$\phi=\Phi-\Phi_c$. We can now compare this expansion to the $\phi^4$
theory in the previous section. 
The redundant term $\phi^3$ can be eliminated, as
usual, by a shift of $\phi$. Comparing with
Eq.~(\ref{Oseries}) we find:
\begin{eqnarray}
 \label{s_tcp}
\Ca\,\hb&=&- (\Delta a + \Delta b \Phi_c^2)\Phi_c\,,\\ \label{p_tcp}
\Ca\,\rb&=& \Delta a  + 3\Delta b \Phi_c^2\,,\\ \label{v_tcp}
\Ca u&=&\frac{20\Phi_c^2}{3}\,,\quad v=\frac{3}{20\Phi_c}\,.
\end{eqnarray}
The $\phi^5$ term causes mixing of $\hb$ and $\rb$ as in
Eq.~(\ref{r_as1}). Using Eqs.~(\ref{h_as1}) and (\ref{r_as1}) to
linear order in $\Delta a$ and $\Delta b$ (i.e., linear order in 
$\Delta T$ and $\Delta\mu$) we find
\begin{align}
\label{eq:hrm-ab1}
 & \Ca\,h =-u^{-1/4}\, \left( \Delta a + \Delta b\Phi_c^2 \right) \Phi_c,\\ \label{eq:hrm-ab2}
 & \Ca\, r = u^{-1/2}\left(\frac{7}{10}\Delta a +\frac{27}{10}\Delta b\Phi_c^2\right). 
\end{align}
Since $a$ and $ b$ are analytic functions of $T$ and
$\mu$ near the critical point we can expand to linear order:
\begin{align}\nonumber
  &\Delta a = a_T \Delta T + a_\mu \Delta \mu \,;\\ \label{eq:abTmu}
  &\Delta b = b_T \Delta T + b_\mu \Delta \mu \,.
\end{align}
Using Eqs.~(\ref{alpha1}),~(\ref{alpha2}), we determine the slopes at the critical point:
\begin{eqnarray} 
  \label{eq:h=0-ab}
 \tan\alpha_1 &=&-\left(\frac{dT}{d\mu}\right)_{h=0}
=  \frac{h_\mu}{h_T} = \frac{a_\mu+b_\mu\Phi_c^2}{a_T+b_T\Phi_c^2}\,;\\
  \label{eq:r=0-ab}
 \tan\alpha_2 &=&-\left(\frac{dT}{d\mu}\right)_{r=0}
=  \frac{r_\mu}{r_T} =  \frac{a_\mu+27b_\mu\Phi_c^2/7}{a_T+27b_T\Phi_c^2/7}\,.
\end{eqnarray}

In general, the two slopes are different and non-universal (i.e.,
depend on the non-universal coefficients $a_\mu$, $a_T$, etc. However,
the limit $m_q\to0$ is special. In this limit the two slopes approach
each other with the difference vanishing as $\Phi_c^2\sim m_q^{2/5}$
(see Eq.~(\ref{eq:phiab})):
\begin{eqnarray}
  \nonumber
   \tan\alpha_1-\tan\alpha_2
=\left(\frac{dT}{d\mu}\right)_{r=0}-\left(\frac{dT}{d\mu}\right)_{h=0}
&=& \frac{20}{7a_T^2} \frac{\partial(a,b)}{\partial(\mu,T)} \Phi_c^2
+ \mathcal O (\Phi_c^4)\\  \label{eq:slope-diff1b}
&=&\frac{20}{7a_T^2} \frac{\partial(a,b)}{\partial(\mu,T)} \left(\frac{3}{8c}\right)^{2/5}m_q^{2/5}
+ \mathcal O (m_q^{4/5})\,,
\end{eqnarray}
where ${\partial(a,b)}/{\partial(\mu,T)} = a_\mu b_T - a_T b_\mu$ is
the Jacobian of the mapping in Eq.~(\ref{eq:abTmu}). 

\begin{figure}[h]
\begin{minipage}{0.5\textwidth}
\begin{center}
\includegraphics[scale=0.49]{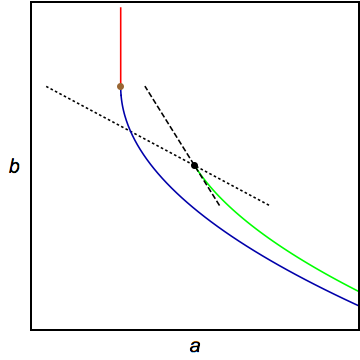}
\end{center}
 \end{minipage}
 \begin{minipage}{0.45\textwidth}
  \begin{center}
\includegraphics[scale=0.36]{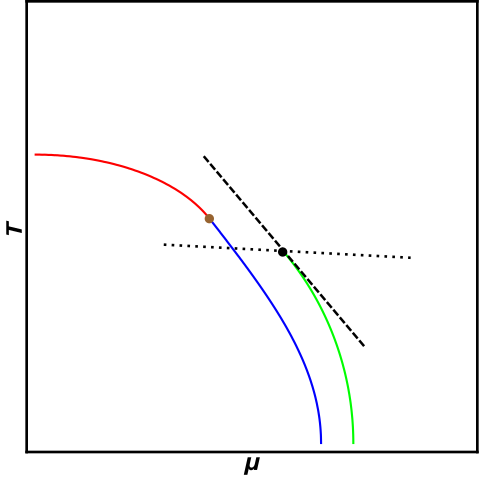}
\end{center}
 \end{minipage}
 \caption{{\it Left:} The phase diagram of the $\Phi^6$
   theory described by Eq.~(\ref{V:-expanded}) in the
   $a-b$ plane. {\it Right:} QCD phase diagram in the $\mu-T$
   plane. The blue and red lines correspond to the first-order and
   second-order phase transitions at $m_q=0$ respectively. They join
   at a tricritical point. The green line represents the first-order
   phase transition at $m_q\neq0$ ending in a critical point. The
   symmetry broken (ordered) phase is in the lower left corner in both
   cases. The slopes of the $h=0$ and $r=0$ lines at the critical
   point are indicated by the dashed and dotted lines respectively.}
\label{fig:top1}
\end{figure}

The relative orientation of the slopes, i.e., the
sign of the slope difference, is determined by the sign of the Jacobian
of the $(a,b)\to(\mu,T)$ mapping. It is positive in the case of the
mapping without reflection and negative otherwise. In that sense, it
is topological. We show how to determine the sign
on Fig.~\ref{fig:top1} by comparing the phase diagram in the vicinity of the
tricritical point in $(a,b)$ coordinates with the standard
scenario of the QCD phase diagram in $(\mu,T)$ coordinates. We see that the
two graphs are topologically the same: the first order transition is
to the right of the tricritical point and the broken (order) phase is
below the tricritical point. This means that the Jacobian of the $(a,b)$
to $(\mu,T)$ is positive (no reflection is involved).
This means that, since $h=0$ slope is negative, the $r=0$ slope must
be less steep, or if $\alpha_1$ itself is small, $\alpha_2$ could be
slightly negative. We shall see in the next Section that in the random
matrix model
both slopes are negative and small (i.e., $\alpha_1>\alpha_2>0$ in the
model).

The Jacobian in Eq.~(\ref{eq:slope-diff3}) can be rewritten in a more geometrically intuitive form in
terms of the difference of slopes of $a=0$ and $b=0$ on the $(\mu,T)$
phase diagram of QCD at $m_q=0$:
\begin{equation}
  \label{eq:Jacobian}
\frac{1}{a_T^2} \frac{\partial(a,b)}{\partial(\mu,T)} 
=
\left(\frac{\partial b}{\partial a}\right)_\mu\left(
\left(\frac{\partial T}{\partial \mu}\right)_{b=0} 
-\left(\frac{\partial T}{\partial \mu}\right)_{a=0} \
\right)\,.
\end{equation}
The $a=0$ slope is, of course, the slope of the chiral phase
transition line at the tricritical point.


One can also determine the dependence of $\rho$ and $w$ on $m_q$ using
Eqs.~(\ref{rho}) and (\ref{w}). Using Eq.~(\ref{V:-expanded}) we find,
in the limit of $m_q\rightarrow0$:
\begin{align}\nonumber
  \sqrt{\Omega^{^2}_{\phi\mu}+\Omega^{^2}_{\phi T}}\sim m_{q}^{1/5},
  \quad
  \sqrt{\Omega^{^2}_{\phi\phi\mu}+\Omega^{^2}_{\phi\phi T}}\sim  m_{q}^{0}, \\
  \label{sc_Omg}
  \Omega_{\phi\mu}\Omega_{\phi\phi T}-\Omega_{\phi\mu}\Omega_{\phi\phi
  T}\sim m_{q}^{3/5},\quad
  \Omega_{\phi\phi\phi\phi}\sim m_{q}^{2/5}\,,
\end{align}
and thus
\begin{equation}\label{eq:rho-w-mq}
\rho\sim m_{q}^{3/10},\qquad w\sim m_{q}^{-1/2}.
\end{equation}

\section{Random Matrix Model}
\label{sec:rmm}

To illustrate the general results derived in the previous section we
consider the random matrix model (RMM) introduced by Halasz {\it et al} in
Ref.~\cite{Halasz:1998qr} in order to describe the chiral symmetry restoring
phase transition in QCD. This is a mean-field model which has features
similar to the effective Landau-Ginsburg potential near a tricritical
point discussed in the previous section.  The QCD pressure
in this model is given by
\begin{equation}
\label{rmm-p}
P(\mu,T,m_q)=-\mathpzc{N}\min_{\phi} \Omega^{RMM}(\Phi;\mu,T,m_q)\,,
\end{equation}
where 
\begin{equation}\label{eq:ORMM}
\Omega^{RMM}(\Phi;\mu,T,m_q)=\Phi^2-\frac{1}{2} \ln \left\{\left[\left(\Phi+m_q\right)^2-\left(\mu+iT\right)^2\right].\left[\left(\Phi+m_q\right)^2-\left(\mu-iT\right)^2\right]\right\}
\end{equation}
and $\mathpzc{N}=n_{\text{inst}}N_f$ where $n_{\text{inst}}\approx 0.5
fm^{-4}$ is the typical instanton number 4-density and $N_f=2$ is the
number of flavors of light quarks. The units for $T,\mu$ and $m_q$
here are such that $T=1$, $\mu=1$ and $m_q=1$ in these units
correspond to approximately $160\,\text{MeV}$, $2300\,\text{MeV}$ and
$100\,\text{MeV}$ respectively (as in Ref.~\cite{Halasz:1998qr}).

To use the results of the previous section we identify
\begin{equation}
  \label{eq:AO=NO}
  \Ca\Omega(\Phi;\mu,T,m_q)=\mathpzc{N}\Omega^{RMM}(2\Phi;\mu,T,m_q)\,,
\end{equation}
which takes into account that $\partial\Omega^{RMM}/\partial m_q = 2\Phi$.

\begin{figure}[h]
	\begin{center}
\includegraphics[height=16em]{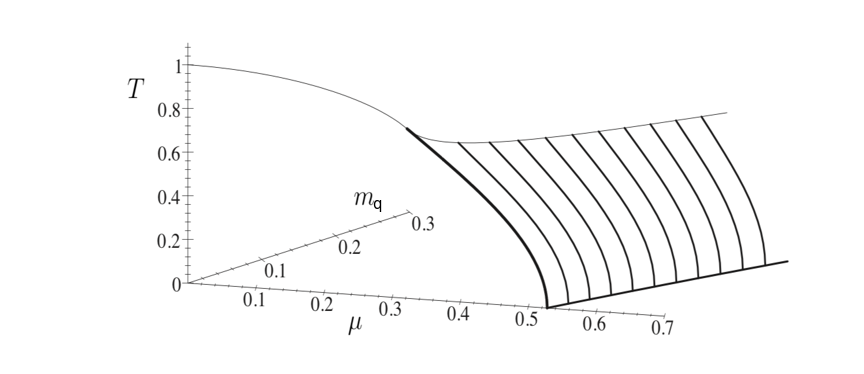}
\caption[]{The phase diagram for the random matrix model in
  Ref.~\cite{Halasz:1998qr}. On the $m_q=0$ plane, the thick and the
  thin lines represent the first-order and the second-order phase
  transitions respectively. Upon turning on $m_q$, the tricritical
  point where these two lines meet turns into a line of Ising-like
  critical points $(\mu_c(m_q),T_c(m_q))$. For the discussion that
  follows, we fix $m_q$ to a particular value and obtain the map from
  $(\mu-\mu_c(m_q),T-T_c(m_q))$ to $(h,r)$ variables.}
\label{fig:pd-3d-rmm}
\end{center}
\end{figure}

The equation of state that follows from this potential, $\partial
\Omega/\partial \Phi=0$, is a fifth order polynomial equation. The
phase diagram resulting from this potential is shown in Fig.~\ref{fig:pd-3d-rmm}.

The tricritical point for this model is at 
 $(\mu_3,T_3)=\left(\sqrt{-1+\sqrt{2}},\sqrt
  {1+\sqrt{2}}\right)/2$. Expanding the potential given by
Eq.~(\ref{eq:ORMM}) we find
	\begin{eqnarray}
	\mathpzc{N^{-1}}\Ca\,\Omega(\Phi;\mu,T)=\mathpzc{N^{-1}}\Ca\,\Omega(0;\mu,T)+\frac{a}{2}\Phi^2+\frac{b}{4}\Phi^4+\frac{c}{6}\Phi^6 -d \Phi+\ldots,
	\end{eqnarray}
	where	\begin{align}
	a= \frac{1}{2}\left(\frac{\mu ^2-T^2}{\left(\mu ^2+T^2\right)^2}+1\right);\quad
	b=\frac{\mu ^4+T^4-6 \mu ^2 T^2}{ 8\left(\mu ^2+T^2\right)^4};\\
	c=\frac{ \left(\mu^2-T ^2\right) \left(\mu ^4+T^4-14 \mu ^2 T^2\right)}{32 \left(\mu ^2+T^2\right)^6};\quad
	d=m_q \frac{T^2-\mu^2}{\left(T^2+\mu^2\right)^2},
	\end{align}		
and dots denote terms such as $\Phi^8$, $m_q\Phi^3$, etc.,
which are of order $a^2$ and smaller, negligible compared to the
terms kept (which are of order $a^{3/2}$), according to the scaling in Eq.~(\ref{eq:3cr-scaling}).

  For a given $m_q$, the critical values $\Phi_c, \mu_c$ and $T_c$ are obtained by simultaneously requiring the first, second and third derivatives of $\Omega$ with respect to $\Phi$ to vanish. As $m_q\rightarrow 0$,
  \begin{equation}
  \mu_c(m_q)=\mu_3+O(m_q^{2/5}) \,, \quad T_c(m_q)=T_3+O(m_q^{2/5}) \,, \quad  \Phi_c(m_q)=\left(6m_q\right)^{1/5}+O(m_q^{3/5})
  \end{equation}
  Using Eq.~(\ref{eq:slope-diff1b}), we can now obtain the slope difference:	
	\begin{eqnarray}\label{rmm_slopedif}
     \tan\alpha_1-\tan\alpha_2
&=&\frac{ 20}{7} (2+\sqrt{2})(6m_q)^{2/5}
+ \mathcal O (m_q^{4/5})\,.
\end{eqnarray}

As $m_{q}\rightarrow 0$, the lines $h=0$ and $r=0$ become nearly
parallel to each other with the difference in their slopes being
proportional to $m_{q}^{2/5}$ as predicted in the previous
section. Comparing Eq.~(\ref{rmm_slopedif}) to
Eq.~(\ref{eq:slope-diff1b}), one can see that
$\partial(a,b)/\partial(\mu,T)$ is positive,  as expected. 


Using more general (finite $m_q$) Eqs.~(\ref{alpha1}-\ref{w},\ref{eq:O''-subs}) we
computed the vales for the parameters
$\alpha_1$, $\alpha_2$, $\rho$ and $w$  at $m_q=0.05$ (which
corresponds to quark masses of 5 MeV in the units of
Ref.~\cite{Halasz:1998qr}) in RMM:
\begin{equation}\label{eq:estimates}
 \alpha_1\sim 13^{\circ},\quad \alpha_2\sim 1 ^{\circ}, \quad \rho\sim 0.5, \quad w\sim 1.4\,.
\end{equation}

\begin{figure}[h]
\begin{minipage}{0.3\textwidth}
\includegraphics[scale=0.4]{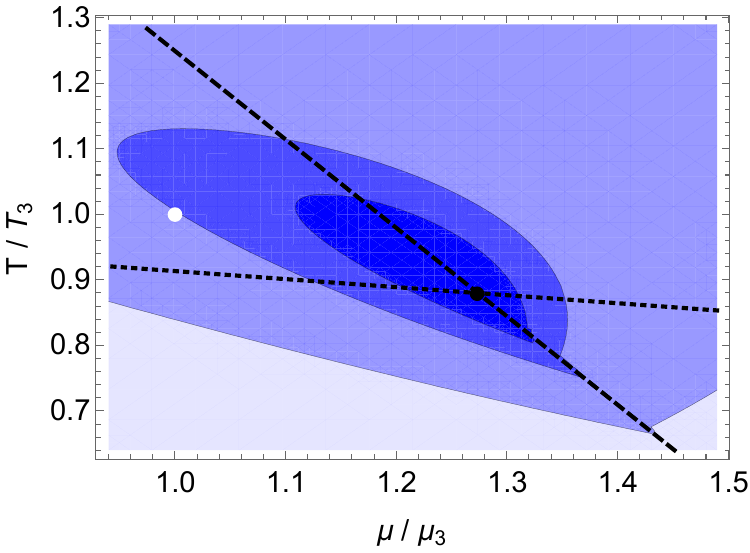}
 \end{minipage}
 \begin{minipage}{0.3\textwidth}
\includegraphics[scale=0.4]{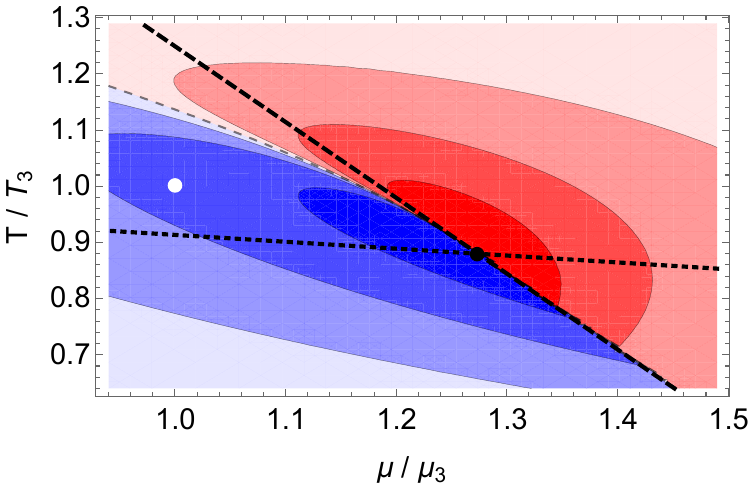}
 \end{minipage}
  \begin{minipage}{0.3\textwidth}
\includegraphics[scale=0.4]{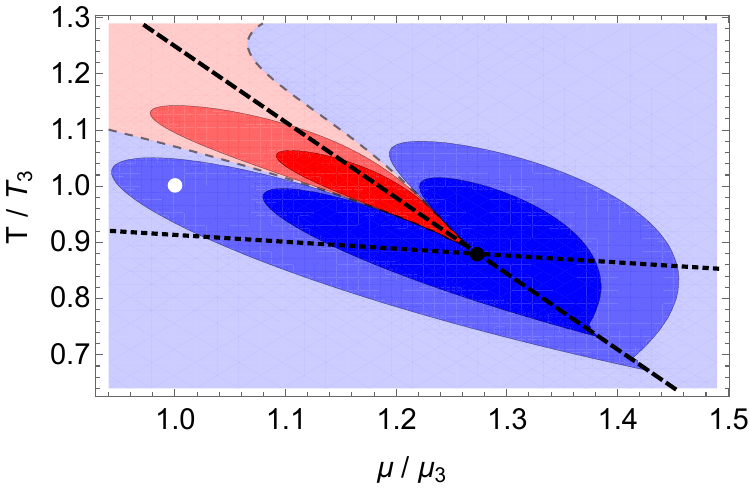}
 \end{minipage}
 \captionof{figure}{Contour plots of susceptibilities
   $\chi_2=P_{\mu\mu}$, $\chi_3=P_{\mu\mu\mu}$ and
   $\chi_4=P_{\mu\mu\mu\mu}$ near the critical point corresponding to
   $m_{q}=0.05$ in the RMM. The black and white dots represent the
   critical point and the tricritical point (at $m_{q}=0$)
   respectively. The dotted and the dashed lines are the $r=0$ and $h=0$
   lines respectively. The slope of $r=0$ is negative for this value of
   $m_{q}$. The negative valued regions are red and positive
   valued regions are blue. Note that the value of $\chi_3$
   along the $h=0$ line on the cross-over side is {\em negative}.}
\label{RMM_chi}
\end{figure}
 
The contour plots of singular pressure derivatives
$\chi_2=P_{\mu\mu}$, $\chi_3=P_{\mu\mu\mu}$ and
$\chi_4=P_{\mu\mu\mu\mu}$ (baryon number cumulants, or
susceptibilities, of second, third and fourth order) around the
critical point at small quark mass are shown in
Fig.~\ref{RMM_chi}. The following observations can be made:
\begin{itemize}
	\item The slopes of $h=0$ and $r=0$ are both negative and
          $h=0$ axis (coexistence line) is steeper than than $r=0$ axis. 
	\item $\rho<w$, which is in qualitative agreement with the
          small $m_q$ scaling in Eqs.~(\ref{eq:rho-w-mq}).
	\item The signs of the cumulants $\chi_2$ and $\chi_4$ on the
          crossover side of $h=0$ line are in agreement with
          Eqs.~(\ref{pTmu}) and~(\ref{pmumuTT}) with
          $g_+''(0)=-1<0$ and $g_+''''(0)=6>0$ according to
          Eq.~(\ref{eq:g'}). 
        \item Most interestingly, the sign of $\chi_3$ on the
          crossover side of $h=0$ line, according to
          Eq.~(\ref{pmumumu1}), is determined by the sign of
          $-r_\mu$. This is clearly seen in Fig.~\ref{RMM_chi}b where
          $\chi_3<0$ in accordance to $r_\mu>0$ ($\alpha_2>0$). If the
          same holds true in QCD, this may have phenomenological
          consequences as the sign of cubic cumulant (skewness) is
          measured in heavy-ion collisions (see also discussion in
          Section~\ref{sec:concl-disc}).
\end{itemize}
 
RMM is a model of QCD, capturing some of its physics, such as
chiral symmetry breaking, and missing other features, such as
confinement. Its results should be treated with caution to avoid
mistaking artifacts for physics. The behavior of the equation of state
near the tricritical point is, however, subject to universality
constraints, which we verified are satisfied by the model. The
numerical values for the mapping parameters we obtained in
Eq.~(\ref{eq:estimates}) should be
treated as estimates, or informed guesses. These parameters are not
universal. However, their dependence on $m_q$ {\em is} universal, and
is manifested in RMM (e.g., the slope difference is small and $\rho<w$
in accordance with Eqs.~(\ref{eq:rho-w-mq})). Since no other
information about these parameters is available as of this writing, we
believe our estimates in Eq.~(\ref{eq:estimates}) could be helpful
for narrowing down the parameter domain of the approximate equations of
state constructed along the lines of Ref.~\cite{Parotto:2018pwx}.

\section{Beyond the mean-field theory}
\label{sec:corrections}

\begin{figure}[h]
  \centering
  \includegraphics[height=8em]{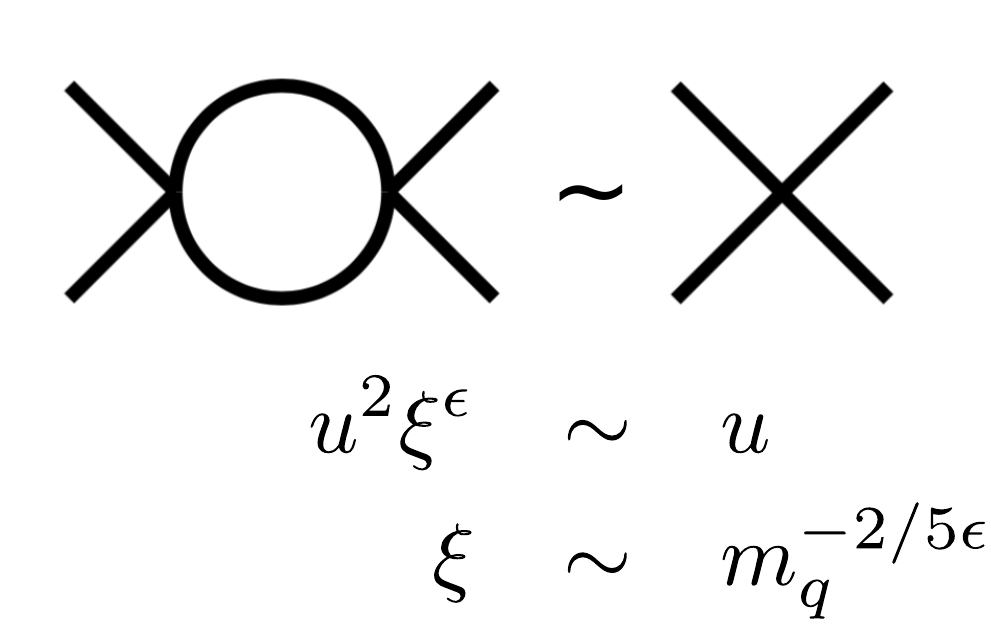}
  \caption{The one-loop contribution of fluctuations compared to the
    tree-level coupling. The fluctuation contribution diverges as
    $\xi^\epsilon$, where $\epsilon=4-d$. The mean-field approximation
  breaks down at sufficiently large $\xi$ when the contribution of
  fluctuations is no longer negligible. The scaling of
$u\sim \Phi_c^2 \sim m_q^{2/5}$ follows from Eq.~(\ref{V:-expanded}).}
  \label{fig:one-loop}
\end{figure}

In Sections~\ref{sec:mean-field-eos}-\ref{sec:rmm}, we discussed the
mean-field theory near a critical point. Within such a theory, we
derived scaling relations for $\tan\alpha_1-\tan\alpha_2$, $\rho$ and
$w$ in the $m_q\rightarrow0$ limit in Eqs.~(\ref{eq:slope-diff1b})
and~(\ref{eq:rho-w-mq}). The mean-field theory should break down
sufficiently close to the critical point in $d=3$ dimensions since the
upper critical dimension near a critical point is $d=4$.  The
breakdown occurs because contribution of fluctuations increases with
increasing correlation length $\xi$ (the fluctuations become coherent
at larger scales). The extent of the region where the mean-field
theory breaks down can be estimated using the Ginsburg criterion by
comparing the strength of the one-loop correction (infrared-divergent
for $d<4$) to the coupling to its tree-level value as shown in
Fig.~\ref{fig:one-loop}.

\begin{figure}[h]
	\begin{center}
\includegraphics[scale=0.5]{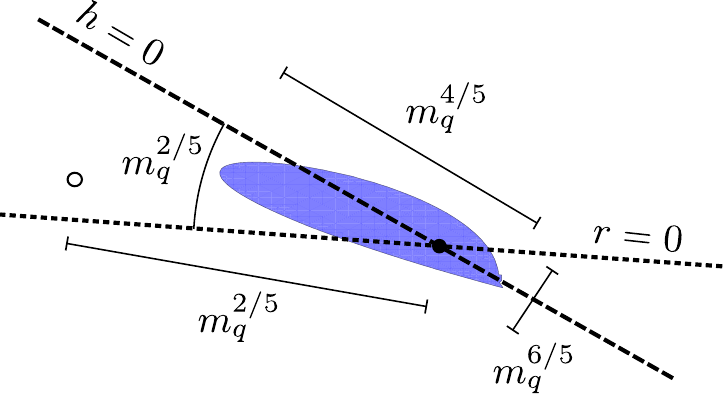}
\end{center}
\caption{Schematic representation of the scaling of various parameters
  characterizing the location, the size and the shape of the Ginzburg
  region (shown in blue) around the QCD critical point in the $T$ vs
  $\mu$ plane for small quark
  mass~$m_q$. The empty circle denotes the location of the tricritical
  point at $m_q=0$. The dotted
  and dashed lines are the $r=0$ and $h=0$ axes, respectively, with an
  angle between them vanishing as $m_q^{2/5}$ in the chiral limit.
}
\label{fig:ginzburg}
	\end{figure}

Since the mean-field limit is essentially weak-coupling limit, a quicker argument is to compare the coupling $u$ expressed in
{\em dimensionless} units, i.e., $u\xi^{\epsilon}$, where $\epsilon=4-d$ is
the mass dimension of $u$, to unity.
Since in the mean-field region $h\sim\xi^{-3}$ and $r\sim \xi^{-2}$,
the boundary of the Ginsburg region where the mean-field theory breaks
down is parametrically given by $h_G\sim m_{q}^{6/5}$,
$r_G\sim m_q^{4/5}$ in $d=3$.  Note that the Ginzburg region is
parametrically small for small $m_{q}$. It is also parametrically
smaller than the distance between the critical and the tricritical
points $b_c\sim m_q^{2/5}$, Eq.~(\ref{eq:phiab}). The characteristic size and
shape of the Ginzburg region is illustrated in
Fig.~\ref{fig:ginzburg}.

In this section we study the effects of the fluctuations to see if and how
our mean-field results are modified in the Ginzburg region.  We are
going to use $\epsilon$ expansion to order $\epsilon^2$ to address
this question. We shall focus on our main result -- the convergence of
the $r=0$ and $h=0$ slopes in the chiral limit $m_q\to0$ described by 
Eq.~(\ref{eq:slope-diff1b}).

The result we derived using mean-field theory could be potentially
modified if the contributions of the fluctuations modify the
expression for $\bar r$ in Eq.~(\ref{p_tcp}). An obvious
contribution to the $\phi^2$ in the effective potential $\Omega$ comes
from a tadpole diagram. This correction, however, does not break the
$Z_2$ symmetry which is necessary to induce the additional mixing of
$r$ and $h$ needed to change the direction of the $r=0$ axis.~\footnote{More explicitly, such
  contributions (infrared singular at the critical point, $r=0$)
  are of order $\epsilon r \log r\,\phi^2$. Together
  with the tree-level term $r\phi^2$, they assemble into
  $r^{\beta\delta-\beta}\phi^2$ as dictated by scaling, where $\beta\delta-\beta = 1 +\mathcal O(\epsilon)$ is the
  actual, non-mean-field value of the corresponding critical
  exponent (see also Ref.~\cite{Wallace:1974,Lawrie_1979}). The correction to
  the critical exponent, obviously, does not
  change the condition $r=0$.}

Therefore, to induce $r-h$ mixing via fluctuations we would need a
$Z_2$ breaking term. Furthermore, $r-h$ mixing violates scaling, since
$h\sim r^{\beta\delta}$ and thus we need terms which violate scaling
by $r^{\beta\delta-1}$. In mean-field theory this
corresponds to scaling violations of order $r^{1/2}$, which are
produced by terms in the potential $\Omega$ which scale as
$r^{5/2}$, i.e., operators of dimension 5. We have already
seen how operator $\phi^5$ induces $r-h$ mixing in
Section~\ref{sec:mean-field-eos}. Here we need to generalize this
discussion to include effects of fluctuations.

As usual, we start at the upper critical dimension $d=4$ and then
expand in $\epsilon=4-d$.  When $\phi$ is a fluctuating field, in
$d=4$, the scaling part of the potential $\Omega$ also includes
additional dimension 4 operator, $(\nabla\phi)^2$, i.e.,
\begin{equation}
  \label{eq:Omega-4}
  \Omega=\frac{1}{2} (\nabla\phi)^2+\frac{\rb}{2} \phi^2+
  \frac{u}{4}\phi^4 -\hb \phi +  \ldots \,,
\end{equation}
where ellipsis denotes higher-dimension operators.  While $\phi^5$ is the
only dimension five $Z_2$ breaking term in the mean-field theory,
when fluctuations of $\phi$ are considered there are two such terms:
$\phi^5$ and $\phi^2\nabla^2\phi$. However, we shall see that only one
special linear combination of these terms has the scaling property
needed to induce $r-h$ mixing when $d<4$.

To identify this linear combination let us observe that using the
transformation of variables $\phi\rightarrow \phi+\Delta\phi$, where
$\Delta\phi=-v(\phi^2-\rb/u)$ similar to Eq.~(\ref{m_as1}), we can
cancel a certain linear combination of $\phi^5$ and
$\phi^2\nabla^2\phi$, while introducing additional $\phi^2$ term:
		\begin{equation}
		\label{eq:Onabla}
		\Delta \Omega=\Delta \phi \frac{\partial
                  \Omega}{\partial \phi}=-v
                \left(u\phi^{5}-\phi^2\nabla^2\phi\right)+v\hb\phi^2+\dots \,,
		\end{equation}
where ellipsis denotes terms which do not affect the mapping (being nonlinear in $\rb$ or simply total derivatives). Therefore, the
effect of the perturbation $vV_3$, where
\begin{equation}
  \label{eq:V3}
 V_3 = u\phi^5 - \phi^2\nabla^2\phi\,,
\end{equation}
is equivalent to the shift $\rb\to \rb+2v\hb$. The correction to scaling
 induced in $G$  due to a perturbation $v_3V_3$ can be absorbed by
``revised scaling''
 \begin{equation}
   \label{eq:Gv}
  G(\rb,\hb) = \rb^{\beta(\delta+1)}\left(
g(\hb\rb^{-\beta\delta}) + v_3 \rb^{\Delta_3} g_3(\hb\rb^{-\beta\delta})
\right) + \ldots
= r^{\beta(\delta+1)} g(hr^{-\beta\delta}) + \ldots\,,
 \end{equation}
where
\begin{equation}
r = \rb + 2v_3 \hb\quad \mbox{and}\quad h=\hb.\label{eq:rhv3}
\end{equation}
This property also guarantees \cite{Nicoll:1981zz,Amit:1984ms}  that operator $V_3$
 is an eigenvector
of the RG matrix of anomalous dimensions which mixes $u\phi^5$ and
$\phi^2\nabla^2\phi$. The corresponding
correction-to-scaling exponent is given by \cite{Nicoll:1981zz,Brezin:1972fc}
\begin{equation}
\Delta_3 = \beta\delta-1=1/2 + \mathcal O(\epsilon^2)\,, \label{eq:D3}
\end{equation}
which is simply the difference between $h$ and $r$ scaling exponents,
as expected, since $V_3$ induces $r-h$ mixing.

\begin{figure}[h]
  \centering
  \includegraphics[height=6em]{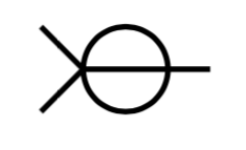}
  \caption{The two-loop diagram responsible for the mixing of
    $\phi^5$ and $\phi^2\nabla^2\phi$ operators.}
  \label{fig:two-loop}
\end{figure}

The other eigenvalue of the anomalous dimension
matrix is
\begin{equation}
\Delta_5=1/2 + \epsilon +\mathcal O(\epsilon^2)\,,\label{eq:D5}
\end{equation}
and the
corresponding eigenvector is
\begin{equation}
 V_5\equiv u\phi^5 - (10 S_5/3)\phi^2\nabla^2\phi.\label{eq:V5}
\end{equation}
The mixing parameter $S_5$ has been calculated in Ref.~\cite{Nicoll:1981zz}:
\begin{equation}
 S_5 = -\epsilon/108 + \mathcal O(u),\label{eq:S5}
\end{equation}
where, consistent with our interest in the $m_q\to0$ limit, we assumed
that $u\ll\epsilon$, since $u\sim m_q^{2/5}$.
The eigenvalue degeneracy is lifted
at one-loop order, however, the mixing only appears at two-loop order
due to the sunset diagram shown in Fig.~\ref{fig:two-loop}. Despite the diagram being of order
$\epsilon^2$, the mixing, i.e., $S_5$, is of order
$\epsilon^2/(\Delta_5-\Delta_3)=\mathcal O(\epsilon)$.

In the case of physical interest, $d=3$, the values of the exponents
$\Delta_3$ and $\Delta_5$ are significantly different. The exponents
$\beta$ and $\delta$ are fairly well known and have been determined using
different methods, including experimental \cite{Guida:1998bx,LeGuillou:1979ixc,Pelissetto:2000ek,ZinnJustin:1989mi}. Correspondingly,
$\Delta_3=\beta\delta-1\approx 0.56$. The exponent $\Delta_5$ is less well
known, but being associated with the leading $Z_2$ asymmetric correction to
scaling, has also been calculated by a variety of methods, such as
functional RG (epsilon expansion estimates also exist, but the
convergence of the epsilon expansion is notoriously poor for this
exponent).  Typically one finds $\Delta_5\approx 1.3-1.6$ \cite{Nicoll:1981zz,Newman:1984zz,Litim:2010tt,Litim:2003kf}. 

The operator $V_5$ does not (and cannot, in $d<4$) change the mixing
of $r$ and $h$ because its scaling dimension, $\Delta_5$ is different
from $\beta\delta-1$.  The corrections to scaling due to operator
$V_5$ show up, as corrections to scaling generally do, in the form:
\begin{equation}
  \label{eq:5}
  G(r,h) =  r^{\beta(\delta+1)}\left(
g(hr^{-\beta\delta}) + v_5 r^{\Delta_5} g_{5}(hr^{-\beta\delta})
\right)\,.
\end{equation}
Since
$\Delta_5>\Delta_3$ the corrections to scaling from $V_5$ are
significantly suppressed compared to the correction accounted
for by revised scaling in Eq.~(\ref{eq:Gv}).

In the purely mean-field theory the operator $\phi^2\nabla^2\phi$ is
essentially zero (there is no spatial dependence) and, therefore, the
coefficient $v_3$ is undefined. In this case, however, we can completely absorb
the $\phi^5$ term by revised scaling as we have described in
Section~\ref{sec:mapping-qcd-3d}. On the other hand, when $\phi$ is a spatially-varying field and its fluctuations are important, we can only absorb the linear combination
$V_3$, and not $V_5$ (in contrast to the mean-field theory where the two
operators are essentially identical and equal $u\phi^5$). The
coefficient $v_3$ of the operator $V_3$ which determines the revised
scaling mixing depends on the coefficients of
the terms $\phi^5$ and $\phi^2\nabla^2\phi$.

Let us denote the
contribution of the operators $\phi^5$ and $\phi^2\nabla^2\phi$ to $\Omega$ in Eq.~(\ref{eq:Omega-4})
as $\Delta \Omega^{A}$, and denote the coefficients of $u\phi^5$,
$\phi^2\nabla^2\phi$ and their linear combinations $V_3$ and $V_5$ so that
\begin{equation}
  \label{eq:w3w5}
  \Delta \Omega^{A} =  w_5 u\phi^5 - w_3\phi^2\nabla^2\phi= v_3 V_3 + v_5 V_5.
\end{equation}

The coefficient $v_3$ responsible for the revised scaling is given by:
\begin{eqnarray}
  \label{eq:V3-w}
  v_3 = (1-10S_5/3)^{-1}\, \left(w_3 - 10S_5 w_5/3\right)\,,
\end{eqnarray}
while $v_5=(1-10S_5/3)^{-1} \left(w_5-w_3\right)$.

For small $m_q$, we have already determined the coefficient of the
$\phi^5$ term (in $d=4$ mean-field theory) by expanding the $\Phi^6$
potential in powers of $\phi=\Phi-\Phi_c$ in
Eq.~(\ref{V:-expanded}), see Eq.~(\ref{v_tcp}):
\begin{equation}
 w_5=\frac{3}{20\Phi_c}\sim m_q^{-1/5}\,.\label{eq:w5}
\end{equation}

 To find the coefficient of the
$\phi^2\nabla^2\phi$ we need to consider fluctuating, i.e., spatially
varying field $\Phi$ and the corresponding potential in
Eq.~(\ref{eq:Omega-4}). For small
$m_q$, the largest contribution to 
$\phi^2\nabla^2\phi$ term comes from the expansion of higher-dimension
term $\Phi^2(\nabla\Phi)^2$, and therefore $w_3$ is vanishing as $\Phi_c\sim m_q^{1/5}$
in the $m_q\to0$ limit.

Hence
\begin{equation}
 w_5\sim m_q^{-1/5}\gg w_3\sim m_q^{1/5}.\label{eq:w5>w3}
\end{equation}
Thus, for $m_q^{2/5}\ll\epsilon\ll1$, the dominant contribution to
$v_3$ in Eq.~(\ref{eq:V3-w}) comes from $w_5$ and, therefore,
 \begin{equation}
v_3= -\frac{S_{5}(\epsilon)}{2\Phi_c}+O(\epsilon^2)\sim\epsilon
m_q^{-1/5}.\label{eq:v3}
\end{equation}
Using Eq.~(\ref{eq:rhv3}) we can now determine the $\mathcal
O(\epsilon)$ correction to the slope difference:
\begin{equation}
  \label{eq:r-change}
\frac{h_\mu}{h_T} - \frac{r_\mu}{r_T} = \frac{\hb_\mu}{\hb_T}- \frac{\rb_\mu+ 2v_3 \hb_\mu}{\rb_T+ 2 v_3 \hb_T}   =
  \left(\frac{\hb_{\mu}}{\hb_T}-\frac{\rb_\mu}{\rb_T}\right)
\left(1+2v_3\frac{\hb_T}{\rb_T}\right)^{-1}\,.
\end{equation}
From Eqs.~(\ref{s_tcp}) and~(\ref{p_tcp}) we conclude that
\begin{equation}\label{eq:slope-diff2b}
\frac{\hb_\mu}{\hb_T} - \frac{\rb_\mu}{\rb_T}
= \frac{2}{a_T^2} \frac{\partial(a,b)}{\partial(\mu,T)} \Phi_c^2
+ \mathcal O (\Phi_c^4)\,,
\end{equation}
and that, to leading
order in $\Phi_c\sim m_q^{1/5}$,
$\hb_T/\rb_T=-\Phi_c$. Substituting into Eq.~(\ref{eq:r-change}) we find
\begin{equation}
\frac{h_\mu}{h_T} - \frac{r_\mu}{r_T} = \frac{2}{a_T^2} \frac{\partial(a,b)}{\partial(\mu,T)} (1+S_5(\epsilon)+O(\epsilon^2)) \Phi_c^{2}
+ \mathcal O (\Phi_c^{4})\,.
 \end{equation}

 We conclude that, at two-loop order, fluctuations do not modify the
 {\em exponent} $m_q^{2/5}$ of the slope difference of $r=0$ and $h=0$ given by
 Eq.(\ref{eq:slope-diff2b}), but change the coefficient by an amount
 $\mathcal O(\epsilon)$.

To summarize,  the leading (and next-to-leading) singular part of QCD pressure can be expressed as
\begin{equation}\label{eq:p-v5}
P_{\text{sing}}(\mu,T)=-Ar^{2-\alpha}\left(
 g(hr^{-\beta\delta})+v_5r^{\Delta_5}g_{5}(hr^{-\beta\delta})
\right)\,,
\end{equation}	
where $h$ and $r$ are given by the map in
Eq.~(\ref{map0}). The leading behavior of the slope
difference of $r=0$ and $h=0$  in the limit of small quark masses is given by
	\begin{eqnarray}
  \label{eq:slope-diff3}
 \tan\alpha_1-\tan\alpha_2 =  \left(\frac{dT}{d\mu}\right)_{r=0} -
   \left(\frac{dT}{d\mu}\right)_{h=0} 
&=&\frac{2}{a_T^2} \frac{\partial(a,b)}{\partial(\mu,T)} \left(\frac{3}{8c}\right)^{2/5}(1+S_5(\epsilon)+O(\epsilon^2)) m_q^{2/5}
+ \mathcal O (m_q^{4/5})
\end{eqnarray}
Note that in the limit $\epsilon=0$ this result does not agree with
Eq.~(\ref{eq:slope-diff1b}) in the mean-field theory. This is because in
this limit $\Delta_5=\Delta_3$ and the second term in
Eq.~(\ref{eq:p-v5}) for pressure can, and should, be absorbed via
revised scaling, modifying the slope of the $r=0$ line (i.e., although
$v_3$ is not well-defined in the mean-field limit, $v_3+v_5=w_5$ is).

Thus, we have verified the robustness of our main result,
$\alpha_1-\alpha_2\sim m_q^{2/5}$, to fluctuation corrections up to
two-loop order. This should not be unexpected
since the scaling $m_q^{2/5}$ is related to the tricritical scaling
exponents ($\delta_t=5$) which are unaffected by fluctuations in
spatial dimension $d=3$ and above.

	\section{Summary and conclusions}
\label{sec:concl-disc}
	
Universality of critical phenomena allows us to predict the leading
singularity of the QCD equation of state near the QCD critical
point. This prediction is expressed in terms of the mapping of the
$(\mu,T)$ variables of QCD onto $(h,r)$ variables of the Ising model,
Eqs.~(\ref{pG}),~(\ref{map0}). The mapping parameters are not dictated by the Ising ($\phi^4$ theory)
universality class and thus far have been treated
as unknown parameters.  In this work we find that, due to the smallness
of quark masses, some of the properties of these parameters are also
universal. This universality is due to the proximity of the
tricritical point.  

Our main focus is on the slope of the $r=0$ line
in the $(\mu,T)$ plane which depends on the amount of the $Z_2$ breaking
at the Ising critical point due to leading corrections to scaling
driven by irrelevant operators, such as $\phi^5$. Our main conclusion
is that in the chiral limit $m_q\to0$, when the critical point of the $\phi^4$
theory approaches the tricritical point of the $\phi^6$ theory, the
$(\mu,T)$/$(h,r)$ mapping becomes singular in a specific way: the difference
between the $r=0$ and $h=0$ slopes vanishes as $m_q^{2/5}$, Eq.~(\ref{eq:slope-diff3}).

The $h=0$ line is essentially the phase coexistence (first-order
transition) line and its slope is negative. Therefore, for sufficiently
small $m_q$, the slope of the $r=0$ line should also become
negative, with the $r=0$ line being less steep than $h=0$ line.

Since the reliable first-principle determination of the critical point
mapping parameters is not available we turn to a model  of QCD -- the random
matrix model. In this model we can see explicitly that for
physical value of the quark mass the $r=0$ slope is indeed negative,
and quite small, $\alpha_2=1^\circ$. We also estimate the values of
mapping parameters $\rho$ and $w$, Eq.~(\ref{eq:estimates}), and find them in
agreement with small $m_q$ scaling expectations from Eq.~(\ref{eq:rho-w-mq}).

The smallness of the slope angle $\alpha_2$ may have
significant consequences for thermodynamic properties near the QCD
critical point. In particular, the magnitude of the baryon cumulants,
determined by the derivatives with respect to the chemical potential at
fixed $T$ should be enhanced. This is because for $\alpha_2=0$
these derivatives are essentially derivatives with respect to $h$,
which are much more singular than $r$ derivatives: e.g.,  $\partial^2
G/\partial h^2\sim r^{-\gamma}$ vs $\partial^2
G/\partial r^2\sim r^{-\alpha}$, where $\gamma\approx 1$ and
$\alpha\ll 1$.

\begin{figure}[h]
	 \begin{minipage}{0.3\textwidth}
    \includegraphics[scale=0.4]{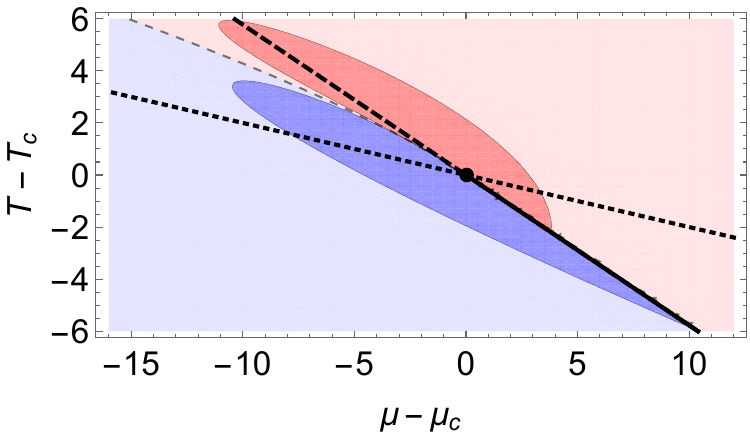}
  \end{minipage}
  \hfill
  \begin{minipage}{0.3\textwidth}
    \includegraphics[scale=0.4]{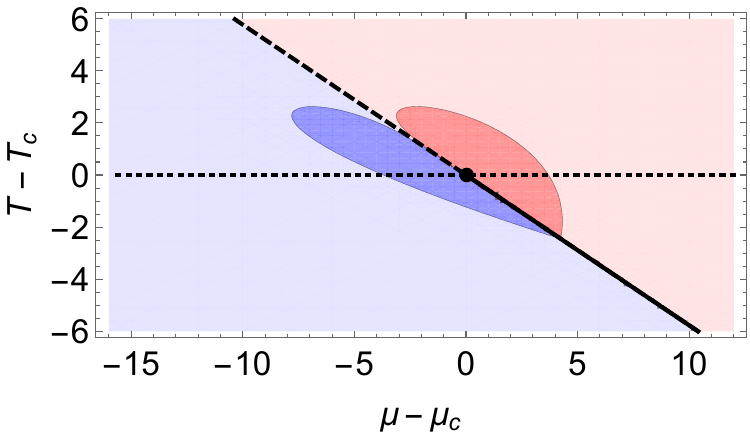}
  \end{minipage}
  \hfill
  \begin{minipage}{0.3\textwidth}
    \includegraphics[scale=0.4]{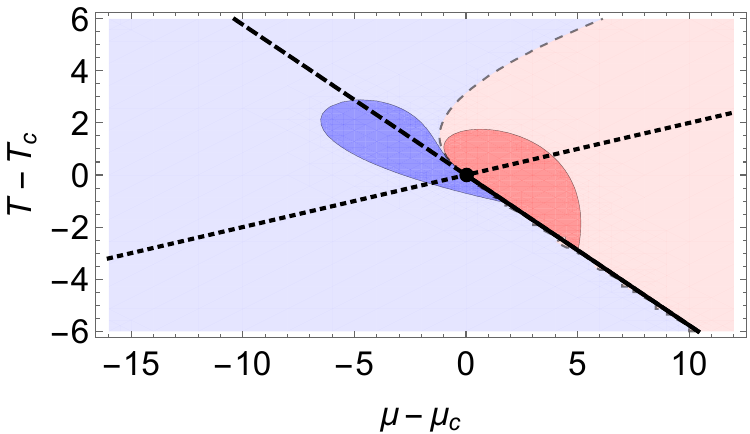}
  \end{minipage}
  \caption{Contours of $\chi_{3}$ when the slope
    (\ref{eq:r-slope}) of the
    $r=0$ line (dotted) is negative, zero and positive (from left
    to right). The contour $\chi_{3}=0$ is shown by the thin dashed
    line. The thick dashed line is the $h=0$ axis (crossover). The
    regions of negative $\chi_3$ are shown in red, and the regions of
    positive value of $\chi_3$ are in blue. Note that $\chi_3$ on the
    crossover line has the same sign as the slope of the $r=0$ line.}
	 \label{ru1}
  \end{figure}

Another interesting conclusion of our study, with potential
phenomenological consequences, is the relation between the sign of the
$r=0$ slope 
\begin{equation}
  \label{eq:r-slope}
  \left(\frac{\partial T}{\partial\mu}\right)_{r=0}
= - \frac{r_\mu}{r_T} = -\tan\alpha_2
\end{equation}
and the sign of the cubic cumulant $\chi_3=P_{\mu\mu\mu}$ of the baryon number (or
skewness) on the crossover line. This relationship can be seen
directly in Eq.~(\ref{pmumumu1}) with $X=\mu$, given $g_+''(0)=-1$, and
is illustrated in Fig.~\ref{ru1} using a $\phi^4$ mean-field
model defined in Eqs.~(\ref{pG}),~(\ref{Oseries1}). Since the 
skewness is measurable in heavy-ion collisions \cite{Adamczyk:2013dal,Luo:2015ewa}, such a measurement
could potentially provide a clue to the values of the nonuniversal
parameters mapping the QCD phase diagram to that of the Ising model.

\acknowledgements	

The authors would like to thank X.~An, G.~Basar, P.~Parotto,
T.~Schaefer and H.-U.~Yee for discussions.  This work is supported by
the U.S. Department of Energy, Office of Science, Office of Nuclear
Physics, within the framework of the Beam Energy Scan Theory (BEST)
Topical Collaboration and grant No. DE-FG0201ER41195.

\appendix

\section{The size of the critical region}
\label{sec:size-critical-region}

Here we describe how the parameters of the mapping control the size of
the critical region. We define the critical region as the region where
the singular part of the equation of state dominates over the regular
part. This comparison cannot be done on the pressure itself, since the
critical contribution to pressure vanishes at the critical point (as
$r^{2-\alpha}$). A reasonable measure of the critical region should be
based on a quantity which is singular at the critical point, such as
the baryon susceptibility, $\chi_2=P_{\mu\mu}$. We shall
evaluate the size of the critical region along the crossover, $h=0$, line. The
singular part of $\chi_2$ is given by, at $h=0$,
\begin{equation}
\label{eqn5}
\chi_2^{\rm sing} \sim AG_{\mu\mu}(r,0)\sim AG_{hh}(r,0) h_{\mu}^2\sim Ar^{-\gamma}\left(\frac{s_1}{wT_c s_{12}}\right)^2\sim A\left(\frac{\Delta \mu}{\rho w T_c c_1}\right)^{-\gamma}\left(\frac{s_1}{wT_c s_{12}}\right)^2
\end{equation}
where $s_1=\sin\alpha_1,\, c_1=\cos\alpha_1$ and
$s_{12}=\sin(\alpha_1-\alpha_2)$. Comparing this to the regular
contribution of order $\chi_2^{\rm reg}\sim T_c^2$, we find for the extent of
the critical region in the $\mu$ direction:
\begin{equation}
\Delta \mu_{\rm CR}\sim T_c\rho w c_1\left(\frac{s_1 \sqrt A}{w
   T_c^2 s_{12}}\right)^{2/\gamma}\label{eq:1}
\end{equation}
Therefore, while increasing parameters $\rho$ and $A$ increases the
size of the critical region, the effect of increasing the parameter
$w$ is the opposite: $\Delta\mu_{\rm CR} \sim w^{1-2/\gamma}$. In the
mean-field theory $\gamma=1$ and $\Delta\mu_{\rm CR}$ is inversely
proportional to~$w$.

 \section{Mapping parameters for the van der Waals equation of state}
\label{sec:vdW}

In this appendix, to illustrate the use of the formalism developed in Section~\ref{sec:mean-field-eos} we shall derive the
equations for the mapping parameters in the van der Waals equation of
state. The well-known equation of state expresses pressure as a
function of particle density $n$ and temperature $T$: 
\begin{equation}\label{eq:P-vdw}
P=\frac{nT}{1-bn}-an^2\,,
\end{equation}
where $a$ and $b$ are van der Waals constants corresponding to the
strength of the particle attraction and the hard-core volume,
respectively. The van der Waals equation of state possesses a
critical point at
\begin{equation}\label{eq:nctcpc}
n_c=\frac1{3b},\quad T_c=\frac{8a}{27 b},\quad P_c= \frac{a}{27b^2}\,.
\end{equation}

The equation of state~(\ref{eq:P-vdw}) can be expressed in the mean-field
(Ginzburg-Landau) form
\begin{equation}
P(\mu,T)=-\Ca\min_{n}\Omega(n,T,\mu)\,,
\end{equation}
where 
\begin{equation}
  \label{eq:OF}
  A\Omega(n,T,\mu) = \mu n - F(T,n)
\end{equation}
is expressed in terms of the free energy $F(n,T)$, which is the
Legendre transform of $P(\mu,T)$:
\begin{eqnarray}\label{eq:F}
F(n,T)&=&n\mu(n,T)-P(\mu(n,T),T)\,.
\end{eqnarray}
In Eq.~(\ref{eq:F}), but not in Eq.~(\ref{eq:OF}), the chemical
potential $\mu(n,T)$ must be determined as a solution to
$n = \partial P/\partial\mu$. This can be done by integrating the
following set of partial differential equations:
\begin{eqnarray}
\label{mu_n}\left(\frac{\partial \mu}{\partial n}\right)_T&=& \frac{1}{n}\left(\frac{\partial p}{\partial n}\right)_T\,;\\
\label{mu_T}\left(\frac{\partial \mu}{\partial T}\right)_n&=&\frac{1}{n}\left(\frac{\partial p}{\partial T}\right)_n-\frac{s}{n}\,;\\
\left(\frac{\partial s}{\partial T}\right)_n&=&\frac{c_v n}{T}\,,
\end{eqnarray}
where $c_v$ is the heat capacity per particle (e.g., $3/2$ for monoatomic gas).
Using the values of $\mu$ and $s$ at the critical point, $\mu_c$ and
$s_c$, as initial conditions one finds
\begin{eqnarray}
\label{mu(n,T)}
\nonumber\mu(n,T)&=&T\left(\log \frac{2bn}{1-bn}-\log
                     \frac{2bn_c}{1-bn_c}\right)
                     +\frac{T}{1-bn}-\frac{T_c}{1-bn_c}\\& &
                                                             -2a(n-n_c) -c_vT\log \frac{T}{T_c}
                                                             +\left(c_v-\frac{s_c}{n_c}\right)(T-T_c)+\mu_c\,.
\end{eqnarray}

Expanding the potential $\Omega$ one obtains
\begin{eqnarray}
\Ca\Omega(n,T,\mu)&=& \Ca\Omega(n_c,T_c,\mu_c) -
                      \left(\Delta\mu-\left(\frac{3}{2}-3 b
                      s_c\right)\Delta T\right)\eta +
                      \frac{27b}{8}\Delta T\eta^2+\frac{9 a
                      b^2}{8}\eta^4-\frac{27a b^3}{40}\eta^5 +
                      \ldots\,,
\end{eqnarray}
where $\eta=n-n_c$, $\Delta T=T-T_c$ and $\Delta\mu = \mu - \mu_c$.
Comparing to Eq.~(\ref{Oseries}) we identify
 \begin{eqnarray}
 \Ca \hb&=&\Delta\mu-\left(\frac{3}{2}-3 b s_c\right)\Delta T,\\
 \Ca \rb&=& \frac{27b}{4}\Delta T,\\
 \Ca u&=&\frac{9 a b^2}{2},\quad
 v=-\frac{3b}{20}.
 \end{eqnarray}
Using Eqs.~(\ref{h_as1}),~(\ref{r_as1}) one then finds
\begin{eqnarray}
 h&=&\Ca^{-3/4}\left(\frac{9ab^2}{2}\right)^{-1/4}\left(\Delta\mu-\left(\frac{3}{2}- 3bs_c\right)\Delta T\right)\\
 r&=&-\Ca^{-1/2}\frac{3}{10}\left(\frac{9a}{2}\right)^{-1/2}\left(\Delta\mu+3\left(b\, s_c-8\right)\Delta T\right)\,.
\end{eqnarray}

Using Eqs.~(\ref{h_sl}),~(\ref{r_sl}),~(\ref{eq:rho-hr})
and~(\ref{eq:w-hr}) one finally obtains
\begin{eqnarray}
\tan\alpha_1&=&-\left(\frac{3}{2}-\frac{s_c}{n_c}\right)^{-1}\,;\\
\tan\alpha_2&=&-\left(24-\frac{s_c}{n_c}\right)^{-1}\,;\\
\rho&=&5 \left(\frac{3P_c}{T_c^4}\right)^{1/4}\sqrt{\frac{4 s_c \left(
        s_c-3n_c\right)+13n_c^2}{s_c \left(
        s_c-48n_c\right)+577n_c^2}}
\,;\\
w&=&\frac{1}{40}\left(\frac{T_c^4}{3P_c}\right)^{3/4} \sqrt{
     \frac{s_c}{n_c} \left( \frac{s_c}{n_c}-48\right)+577}\,.
\end{eqnarray}

 \newpage
\mbox{}\vskip -3.5em
\bibliography{ref1}

\providecommand{\href}[2]{#2}\begingroup\raggedright\begin{thebibliography}{10}

\bibitem{Stephanov:2004wx}
M.~A. Stephanov, ``{QCD phase diagram and the critical point},''
  \href{http://dx.doi.org/10.1142/S0217751X05027965}{{\em Prog. Theor. Phys.
  Suppl.} {\bfseries 153} (2004) 139--156},
  \href{http://arxiv.org/abs/hep-ph/0402115}{{\ttfamily arXiv:hep-ph/0402115
  [hep-ph]}}.
[Int. J. Mod. Phys.A20,4387(2005)].

\bibitem{Luo:2017faz}
X.~Luo and N.~Xu, ``{Search for the QCD Critical Point with Fluctuations of
  Conserved Quantities in Relativistic Heavy-Ion Collisions at RHIC : An
  Overview},'' \href{http://dx.doi.org/10.1007/s41365-017-0257-0}{{\em Nucl.
  Sci. Tech.} {\bfseries 28} no.~8, (2017) 112},
\href{http://arxiv.org/abs/1701.02105}{{\ttfamily arXiv:1701.02105 [nucl-ex]}}.

\bibitem{Ding:2015ona}
H.-T. Ding, F.~Karsch, and S.~Mukherjee, ``{Thermodynamics of
  strong-interaction matter from Lattice QCD},''
  \href{http://dx.doi.org/10.1142/S0218301315300076}{{\em Int. J. Mod. Phys.}
  {\bfseries E24} no.~10, (2015) 1530007},
\href{http://arxiv.org/abs/1504.05274}{{\ttfamily arXiv:1504.05274 [hep-lat]}}.

\bibitem{Stephanov:1998dy}
M.~A. Stephanov, K.~Rajagopal, and E.~V. Shuryak, ``{Signatures of the
  tricritical point in QCD},''
  \href{http://dx.doi.org/10.1103/PhysRevLett.81.4816}{{\em Phys. Rev. Lett.}
  {\bfseries 81} (1998) 4816--4819},
\href{http://arxiv.org/abs/hep-ph/9806219}{{\ttfamily arXiv:hep-ph/9806219
  [hep-ph]}}.

\bibitem{Rajagopal:1992qz}
K.~Rajagopal and F.~Wilczek, ``{Static and dynamic critical phenomena at a
  second order QCD phase transition},''
  \href{http://dx.doi.org/10.1016/0550-3213(93)90502-G}{{\em Nucl. Phys.}
  {\bfseries B399} (1993) 395--425},
\href{http://arxiv.org/abs/hep-ph/9210253}{{\ttfamily arXiv:hep-ph/9210253
  [hep-ph]}}.

\bibitem{Gavin:1993yk}
S.~Gavin, A.~Gocksch, and R.~D. Pisarski, ``{QCD and the chiral critical
  point},'' \href{http://dx.doi.org/10.1103/PhysRevD.49.R3079}{{\em Phys. Rev.}
  {\bfseries D49} (1994) R3079--R3082},
\href{http://arxiv.org/abs/hep-ph/9311350}{{\ttfamily arXiv:hep-ph/9311350
  [hep-ph]}}.

\bibitem{Widom}
B.~Widom, ``Equation of state in the neighborhood of the critical point,''
  \href{http://dx.doi.org/10.1063/1.1696618}{{\em The Journal of Chemical
  Physics} {\bfseries 43} no.~11, (1965) 3898--3905}.

\bibitem{Rehr:1973zz}
J.~J. Rehr and N.~D. Mermin, ``{Revised Scaling Equation of State at the
  Liquid-Vapor Critical Point},''
\href{http://dx.doi.org/10.1103/PhysRevA.8.472}{{\em Phys. Rev.} {\bfseries A8}
  (1973) 472--480}.

\bibitem{Karsch:2001nf}
F.~Karsch, E.~Laermann, and C.~Schmidt, ``{The Chiral critical point in
  three-flavor QCD},''
  \href{http://dx.doi.org/10.1016/S0370-2693(01)01114-5}{{\em Phys. Lett.}
  {\bfseries B520} (2001) 41--49},
\href{http://arxiv.org/abs/hep-lat/0107020}{{\ttfamily arXiv:hep-lat/0107020
  [hep-lat]}}.

\bibitem{Hatta:2002sj}
Y.~Hatta and T.~Ikeda, ``{Universality, the QCD critical / tricritical point
  and the quark number susceptibility},''
  \href{http://dx.doi.org/10.1103/PhysRevD.67.014028}{{\em Phys. Rev.}
  {\bfseries D67} (2003) 014028},
\href{http://arxiv.org/abs/hep-ph/0210284}{{\ttfamily arXiv:hep-ph/0210284
  [hep-ph]}}.

\bibitem{Nonaka:2004pg}
C.~Nonaka and M.~Asakawa, ``{Hydrodynamical evolution near the QCD critical end
  point},'' \href{http://dx.doi.org/10.1103/PhysRevC.71.044904}{{\em Phys.
  Rev.} {\bfseries C71} (2005) 044904},
\href{http://arxiv.org/abs/nucl-th/0410078}{{\ttfamily arXiv:nucl-th/0410078
  [nucl-th]}}.

\bibitem{Bluhm:2006av}
M.~Bluhm and B.~Kampfer, ``{Quasi-particle perspective on critical
  end-point},'' \href{http://dx.doi.org/10.22323/1.029.0004}{{\em PoS}
  {\bfseries CPOD2006} (2006) 004},
\href{http://arxiv.org/abs/hep-ph/0611083}{{\ttfamily arXiv:hep-ph/0611083
  [hep-ph]}}.

\bibitem{Parotto:2018pwx}
P.~Parotto, M.~Bluhm, D.~Mroczek, M.~Nahrgang, J.~Noronha-Hostler,
  K.~Rajagopal, C.~Ratti, T.~Schäfer, and M.~Stephanov, ``{Lattice-QCD-based
  equation of state with a critical point},''
\href{http://arxiv.org/abs/1805.05249}{{\ttfamily arXiv:1805.05249 [hep-ph]}}.

\bibitem{Akamatsu:2018vjr}
Y.~Akamatsu, D.~Teaney, F.~Yan, and Y.~Yin, ``{Transits of the QCD Critical
  Point},''
\href{http://arxiv.org/abs/1811.05081}{{\ttfamily arXiv:1811.05081 [nucl-th]}}.

\bibitem{Rummukainen:1998as}
K.~Rummukainen, M.~Tsypin, K.~Kajantie, M.~Laine, and M.~E. Shaposhnikov,
  ``{The Universality class of the electroweak theory},''
  \href{http://dx.doi.org/10.1016/S0550-3213(98)00494-5}{{\em Nucl. Phys.}
  {\bfseries B532} (1998) 283--314},
\href{http://arxiv.org/abs/hep-lat/9805013}{{\ttfamily arXiv:hep-lat/9805013
  [hep-lat]}}.

\bibitem{Lee:1952ig}
T.~D. Lee and C.-N. Yang, ``{Statistical theory of equations of state and phase
  transitions. 2. Lattice gas and Ising model},''
  \href{http://dx.doi.org/10.1103/PhysRev.87.410}{{\em Phys. Rev.} {\bfseries
  87} (1952) 410--419}.
[,157(1952)].

\bibitem{Fisher:1978pf}
M.~E. Fisher, ``{Yang-Lee edge singularity and $\phi^3$ field theory},''
\href{http://dx.doi.org/10.1103/PhysRevLett.40.1610}{{\em Phys.\ Rev.\ Lett.}
  {\bfseries 40} (1978) 1610}.

\bibitem{Fonseca:2001dc}
P.~Fonseca and A.~Zamolodchikov, ``{Ising field theory in a magnetic field:
  Analytic properties of the free energy},''
\href{http://arxiv.org/abs/hep-th/0112167}{{\ttfamily arXiv:hep-th/0112167
  [hep-th]}}.

\bibitem{An:2017brc}
X.~An, D.~Mesterházy, and M.~A. Stephanov, ``{On spinodal points and Lee-Yang
  edge singularities},'' \href{http://dx.doi.org/10.1088/1742-5468/aaac4a}{{\em
  J. Stat. Mech.} {\bfseries 1803} no.~3, (2018) 033207},
\href{http://arxiv.org/abs/1707.06447}{{\ttfamily arXiv:1707.06447 [hep-th]}}.

\bibitem{Halasz:1998qr}
A.~M. Halasz, A.~D. Jackson, R.~E. Shrock, M.~A. Stephanov, and J.~J.~M.
  Verbaarschot, ``{On the phase diagram of QCD},''
  \href{http://dx.doi.org/10.1103/PhysRevD.58.096007}{{\em Phys. Rev.}
  {\bfseries D58} (1998) 096007},
\href{http://arxiv.org/abs/hep-ph/9804290}{{\ttfamily arXiv:hep-ph/9804290
  [hep-ph]}}.

\bibitem{Wallace:1974}
D.~J. Wallace and R.~K.~P. Zia, ``{Parametric models and the Ising equation of
  state at order $\varepsilon^3$},''
  \href{http://dx.doi.org/10.1088/0022-3719/7/19/008}{{\em J.\ Phys.\ C}
  {\bfseries 7} (1974) 3480}.

\bibitem{Lawrie_1979}
I.~D. Lawrie, ``Tricritical scaling and renormalisation of $\phi$6operators in
  scalar systems near four dimensions,''
  \href{http://dx.doi.org/10.1088/0305-4470/12/6/023}{{\em Journal of Physics
  A: Mathematical and General} {\bfseries 12} no.~6, (Jun, 1979) 919--940}.

\bibitem{Nicoll:1981zz}
J.~F. Nicoll, ``{Critical phenomena of fluids: Asymmetric
  Landau-Ginzburg-Wilson model},''
\href{http://dx.doi.org/10.1103/PhysRevA.24.2203}{{\em Phys. Rev.} {\bfseries
  A24} (1981) 2203--2220}.

\bibitem{Amit:1984ms}
D.~J. Amit, {\em {Field theory, the renormalization group, and critical
  phenomena}}.
\newblock Singapore, World Scientific,
1984.
\newblock

\bibitem{Brezin:1972fc}
E.~Brezin, D.~J. Wallace, and K.~G. Wilson, ``{Feynman graph expansion for the
  equation of state near the critical point (Ising-like case)},''
\href{http://dx.doi.org/10.1103/PhysRevLett.29.591}{{\em Phys. Rev. Lett.}
  {\bfseries 29} (1972) 591--594}.

\bibitem{Guida:1998bx}
R.~Guida and J.~Zinn-Justin, ``{Critical exponents of the N vector model},''
  \href{http://dx.doi.org/10.1088/0305-4470/31/40/006}{{\em J. Phys.}
  {\bfseries A31} (1998) 8103--8121},
\href{http://arxiv.org/abs/cond-mat/9803240}{{\ttfamily arXiv:cond-mat/9803240
  [cond-mat]}}.

\bibitem{LeGuillou:1979ixc}
J.~C. Le~Guillou and J.~Zinn-Justin, ``{Critical Exponents from Field
  Theory},''
\href{http://dx.doi.org/10.1103/PhysRevB.21.3976}{{\em Phys. Rev.} {\bfseries
  B21} (1980) 3976--3998}.

\bibitem{Pelissetto:2000ek}
A.~Pelissetto and E.~Vicari, ``{Critical phenomena and renormalization group
  theory},'' \href{http://dx.doi.org/10.1016/S0370-1573(02)00219-3}{{\em Phys.
  Rept.} {\bfseries 368} (2002) 549--727},
\href{http://arxiv.org/abs/cond-mat/0012164}{{\ttfamily arXiv:cond-mat/0012164
  [cond-mat]}}.

\bibitem{ZinnJustin:1989mi}
J.~Zinn-Justin, ``{Quantum field theory and critical phenomena},''
{\em Int. Ser. Monogr. Phys.} {\bfseries 77} (1989) 1--914.

\bibitem{Newman:1984zz}
K.~E. Newman and E.~K. Riedel, ``{Critical exponents by the scaling-field
  method: The isotropic N-vector model in three dimensions},''
\href{http://dx.doi.org/10.1103/PhysRevB.30.6615}{{\em Phys. Rev.} {\bfseries
  B30} (1984) 6615--6638}.

\bibitem{Litim:2010tt}
D.~F. Litim and D.~Zappala, ``{Ising exponents from the functional
  renormalisation group},''
  \href{http://dx.doi.org/10.1103/PhysRevD.83.085009}{{\em Phys. Rev.}
  {\bfseries D83} (2011) 085009},
\href{http://arxiv.org/abs/1009.1948}{{\ttfamily arXiv:1009.1948 [hep-th]}}.

\bibitem{Litim:2003kf}
D.~F. Litim and L.~Vergara, ``{Subleading critical exponents from the
  renormalization group},''
  \href{http://dx.doi.org/10.1016/j.physletb.2003.11.047}{{\em Phys. Lett.}
  {\bfseries B581} (2004) 263--269},
\href{http://arxiv.org/abs/hep-th/0310101}{{\ttfamily arXiv:hep-th/0310101
  [hep-th]}}.

\bibitem{Adamczyk:2013dal}
{\bfseries STAR} Collaboration, L.~Adamczyk {\em et~al.}, ``{Energy Dependence
  of Moments of Net-proton Multiplicity Distributions at RHIC},''
  \href{http://dx.doi.org/10.1103/PhysRevLett.112.032302}{{\em Phys. Rev.
  Lett.} {\bfseries 112} (2014) 032302},
\href{http://arxiv.org/abs/1309.5681}{{\ttfamily arXiv:1309.5681 [nucl-ex]}}.

\bibitem{Luo:2015ewa}
{\bfseries STAR} Collaboration, X.~Luo, ``{Energy Dependence of Moments of
  Net-Proton and Net-Charge Multiplicity Distributions at STAR},'' {\em PoS}
  {\bfseries CPOD2014} (2015) 019,
\href{http://arxiv.org/abs/1503.02558}{{\ttfamily arXiv:1503.02558 [nucl-ex]}}.

\end{thebibliography}\endgroup
 \bibliographystyle{utphys}

\end{document}